\documentclass[twocolumn]{aastex631}

\usepackage{xcolor}
\usepackage{hyperref}

\makeatletter 
  \patchcmd{\NAT@citex}
    {\@citea\NAT@hyper@{%
      \NAT@nmfmt{\NAT@nm}%
      \hyper@natlinkbreak{\NAT@aysep\NAT@spacechar}{\@citeb\@extra@b@citeb}%
      \NAT@date}}
    {\@citea\NAT@nmfmt{\NAT@nm}%
    \NAT@aysep\NAT@spacechar\NAT@hyper@{\NAT@date}}{}{}

  \patchcmd{\NAT@citex}
    {\@citea\NAT@hyper@{%
      \NAT@nmfmt{\NAT@nm}%
      \hyper@natlinkbreak{\NAT@spacechar\NAT@@open\if*#1*\else#1\NAT@spacechar\fi}%
        {\@citeb\@extra@b@citeb}%
      \NAT@date}}
    {\@citea\NAT@nmfmt{\NAT@nm}%
    \NAT@spacechar\NAT@@open\if*#1*\else#1\NAT@spacechar\fi\NAT@hyper@{\NAT@date}}
    {}{}
\makeatother

\newcommand{\lya}{Ly$\alpha$}
\newcommand{\ha}{H$\alpha$}
\newcommand{\hb}{H$\beta$}

\newcommand{\hi}{H\textsc{i}}

\begin{document}

\title{The Lyman-alpha and Continuum Origins Survey. I. Survey description and Ly$\alpha$ imaging}

\author[0000-0003-1767-6421]{Alexandra Le Reste}
\affiliation{Minnesota Institute for Astrophysics, University of Minnesota, 116 Church Street SE, Minneapolis, MN 55455, USA}

\author[0000-0002-9136-8876]{Claudia Scarlata}
\affiliation{Minnesota Institute for Astrophysics, University of Minnesota, 116 Church Street SE, Minneapolis, MN 55455, USA}

\author[0000-0001-8587-218X]{Matthew J. Hayes}
\affiliation{Department of Astronomy, Oskar Klein Centre, Stockholm University,106 91 Stockholm, Sweden}

\author[0000-0003-0470-8754]{Jens Melinder}
\affiliation{Department of Astronomy, Oskar Klein Centre, Stockholm University,106 91 Stockholm, Sweden}

\author[0000-0001-8419-3062]{Alberto Saldana-Lopez}
\affiliation{Department of Astronomy, Oskar Klein Centre, Stockholm University,106 91 Stockholm, Sweden}

\author[0000-0002-2838-9033]{Aaron Smith}
\affiliation{Department of Physics, The University of Texas at Dallas, Richardson, TX 75080, USA}

\author[0000-0002-1025-7569]{Axel Runnholm}
\affiliation{Department of Astronomy, Oskar Klein Centre, Stockholm University,106 91 Stockholm, Sweden}

\author[0000-0001-8792-3091]{Yu-Heng Lin}
\affiliation{Caltech/IPAC, 1200 E. California Blvd. Pasadena, CA 91125, USA}

\author[0000-0001-5758-1000]{Ricardo O. Amor\'{i}n}
\affiliation{Instituto de Astrof\'{i}sica de Andaluc\'{i}a (CSIC), Apartado 3004, 18080 Granada, Spain}

\author[0000-0002-7570-0824]{Hakim Atek}
\affiliation{Institut d'Astrophysique de Paris, CNRS, Sorbonne Université, 98bis Boulevard Arago, 75014, Paris, France}

\author[0000-0002-2724-8298]{Sanchayeeta Borthakur}
\affiliation{School of Earth and Space Exploration, Arizona State University, 781 Terrace Mall, Tempe, AZ 85287, USA}

\author[0000-0003-4166-2855]{Cody A. Carr}
 \affiliation{Center for Cosmology and Computational Astrophysics, Institute for Advanced Study in Physics \\ Zhejiang University, Hangzhou 310058,  China}
\affiliation{Institute of Astronomy, School of Physics, Zhejiang University, Hangzhou 310058,  China}

\author{Brian Fleming}
\affiliation{Laboratory for Atmospheric and Space Physics, University of Colorado, 3665 Discovery Dr., Boulder, CO 80303}

\author[0000-0002-0159-2613]{Sophia R. Flury}
\affiliation{Institute for Astronomy, University of Edinburgh, Royal Observatory, Edinburgh, EH9 3HJ, UK}

\author[0000-0002-7831-8751]{Mauro Giavalisco}
\affiliation{University of Massachusetts Amherst, 710 North Pleasant Street, Amherst, MA 01003-9305, USA}

\author{Alaina Henry}
\affiliation{Center for Astrophysical Sciences, Department of Physics \& Astronomy, Johns Hopkins University, Baltimore, MD 21218, USA}
\affiliation{Space Telescope Science Institute; 3700 San Martin Drive, Baltimore, MD 21218, USA}

\author{Anne E. Jaskot}
\affiliation{Department of Astronomy, Williams College, Williamstown, MA 01267, USA}

\author[0000-0001-7673-2257]{Zhiyuan Ji}
\affiliation{Steward Observatory, University of Arizona, 933 N. Cherry Avenue, Tucson, AZ 85721, USA}

\author[0000-0003-1187-4240]{Intae Jung}
\affiliation{Space Telescope Science Institute, 3700 San Martin Drive Baltimore, MD 21218, United States}

\author[0000-0002-6085-5073]{Floriane Leclercq}
\affiliation{CNRS, Centre de Recherche Astrophysique de Lyon UMR5574, Univ Lyon, Univ Lyon1, Ens de Lyon, F-69230 Saint-Genis-Laval, France}

\author[0000-0001-8442-1846]{Rui Marques-Chaves}
\affiliation{Observatoire de Gen\`eve, Universit\'e de Gen\`eve, Chemin Pegasi 51, 1290 Versoix, Switzerland}

\author[0000-0003-0503-4667]{Stephan R. McCandliss}
\affiliation{Johns Hopkins University, Department of Physics \& Astronomy, Center for Astrophysical Sciences, 3400 North Charles Street, Baltimore, MD, USA, 21218
}

\author[0000-0002-5808-1320]{M. S. Oey}
\affiliation{Astronomy Department, University of Michigan, 1085 South University Avenue, Ann Arbor, MI 48109, USA}

\author[0000-0002-3005-1349]{G\"{o}ran \"{O}stlin}
\affiliation{Department of Astronomy, Oskar Klein Centre, Stockholm University,106 91 Stockholm, Sweden}

\author[0000-0002-5269-6527]{Swara Ravindranath }
\affiliation{Astrophysics Science Division, NASA Goddard Space Flight Center, 8800 Greenbelt Road, Greenbelt, MD 20771, USA}
\affiliation{Center for Research and Exploration in Space Science and Technology II, Department of Physics, Catholic University of America, 620 Michigan Avenue N.E., Washington, DC 20064, USA}

\author[0000-0001-7144-7182]{Daniel Schaerer}
\affiliation{Observatoire de Gen\`eve, Universit\'e de Gen\`eve, Chemin Pegasi 51, 1290 Versoix, Switzerland}
\affiliation{CNRS, IRAP, 14 Avenue E. Belin, 31400 Toulouse, France}

\author[0000-0001-5331-2030]{Trinh X. Thuan}
\affiliation{Astronomy Department, University of Virginia, P.O. Box 400325, Charlottesville, VA 22904-4325,USA}

\author[0000-0002-9217-7051]{Xinfeng Xu}
\affiliation{Department of Physics and Astronomy, Northwestern University,2145 Sheridan Road, Evanston, IL, 60208, USA.}
\affiliation{Center for Interdisciplinary Exploration and Research in Astrophysics (CIERA), 1800 Sherman Avenue, Evanston, IL, 60201, USA.}

\begin{abstract}
Understanding the mechanisms driving the escape of ionizing or Lyman continuum (LyC) emission from the interstellar medium of galaxies is necessary to constrain the evolution of reionization, and the sources responsible for it. While progress has been made in identifying the global galaxy properties linked to the ionizing escape fraction $f_{\rm esc}^{\rm LyC}$, little is currently known about how spatially resolved galaxy properties impact it. We present Hubble Space Telescope (HST) imaging data obtained in the Lyman $\alpha$ and Continuum Origins Survey (LaCOS). LaCOS consists of HST imaging covering rest-frame optical and UV bands for a subsample of 42 galaxies in the Low redshift Lyman Continuum Survey (LzLCS), including 22 LyC emitters ($f_{\rm esc}^{\rm LyC}=0.01-0.49$). Here, we describe the sample, observations and data reduction, and investigate connections between global and sub-kpc Lyman $\alpha$ (\lya) emission, and $f_{\rm esc}^{\rm LyC}$. We confirm the correlation between $f_{\rm esc}^{\rm LyC}$ and EW$_{\rm Ly\alpha}$, and the anticorrelation with $r_{50}$, when using values obtained via global photometry. We also find correlations previously found with spectroscopy with global photometric $L_{\rm Ly\alpha}$, $f_{\rm esc}^{\rm Ly\alpha}$, $\Sigma_{\rm SFR}$, and $f_{\rm esc}^{\rm LyC}$, but with a smaller degree of correlations ($\overline{|\Delta\tau|}\sim0.1$). We find correlations are strongest between \lya\ observables ($L_{\rm Ly\alpha}$, EW$_{\rm Ly\alpha}$) and $f_{\rm esc}^{\rm LyC}$ when measured in a small aperture around the brightest UV source in each galaxy. We interpret these results as evidence that LyC photons escaping on the line-of-sight are contributed by a small number of UV-bright compact regions in most galaxies in LaCOS.
\end{abstract}


\section{Introduction} 
\label{sec:intro}

The Epoch of Reionization (EoR) is a key cosmological period at $z \geq 5.5$  \citep{Becker2001,Fan2006,Bosman2022} during which the first sources of light almost completely ionize hydrogen in the intergalactic medium (IGM). The onset and progress of reionization largely depends on the number and ionizing efficiency of sources at this epoch emitting in the Lyman continuum range (LyC; with $h\nu>13.6\, \rm{eV}$ or $\lambda<912\,$\AA).
The globally-averaged ionization rate density $\dot{n}_{\rm ion}$ is the product of three parameters: the UV luminosity density function $\rho_{\rm{UV}}$, the ionizing photons production efficiency $\xi_{\rm ion}$ and the escape fraction of ionizing photons $f_{\rm esc}^{\rm LyC}$ \citep[e.g.][]{Robertson2022}.
These parameters not only offer insights into the evolution of the last major phase transition in the Universe, but also encode important information about the first sources of light and the physics of the interstellar and circumgalactic media in the early Universe. As a result, significant effort has been made in both observational and computational astrophysics in the past decades to constrain these three parameters for different UV-emitting sources. 

Observations of galaxies into the EoR with the Hubble Space Telescope (HST) and the James Webb Space Telescope (JWST) have yielded improved estimations of the UV luminosity function \citep{Harikane2023,Finkelstein2023,Perez-Gonzalez2023,Adams2024,Finkelstein2024,Bagley2024,McLeod2024,Donnan2024} and ionizing photon production efficiency \citep{Atek2024,Simmonds2024,Harshan2024,Mascia2024,Llerena2024,Hayes2025} of the galaxy population at redshift $z>9$. These observations have found a larger abundance of UV-bright galaxies than previously thought, and an increased ionizing production efficiency for UV-faint, low-mass sources. Despite these advances, a full picture of reionization history requires constraints on the escape fraction. Yet, measurements of $f_{\rm esc}^{\rm LyC}$ during the EoR are impeded by the fraction of remaining neutral hydrogen in the IGM, a small amount of which is nevertheless sufficient to absorb all LyC photons that escape from the galaxies where they were produced \citep{Inoue2014}. As a result, the escape fraction cannot be constrained directly at this epoch, leaving uncertainties on the physical mechanisms enabling galaxies to reionize the Universe. Changes of the $f_{\rm esc}^{\rm LyC}$ dependence on observed galaxy properties can significantly alter the reionization history \citep[e.g.][]{Finkelstein2019,Naidu2020,Lin2024}. Thus, determining the value of this parameter and how it varies across time and galaxy populations is vital to understanding the EoR.

To circumvent limitations posed by the IGM absorption in the early Universe, observations have turned to star-forming galaxies at lower redshift ($z<4$) to search for LyC-emitting galaxies, and ultimately identify the properties that facilitate LyC escape \citep{Bergvall2006,Leitet2013,Borthakur2014,Mostardi2015,deBarros2016,Vanzella2016,Shapley2016,Izotov2016a,Izotov2016b,Izotov2018a,Izotov2018b,Ji2020,Izotov2021,Izotov2022,Rutkowski2017,Rivera-Thorsen2019,Wang2019,Saxena2022,Flury2022a,Xu2022,Roy2024,Citro2024}. These surveys offer the potential to explore and evaluate the galaxy properties linked to ionizing radiation escape. Low-redshift surveys have shown that the detection of LyC emission is generally associated with galaxy properties characteristic of young, intensely star-forming, compact and low-metallicity galaxies \citep[e.g.][]{Izotov2016b,Flury2022b,Flury2024,Carr2024}, although see \citet{Roy2024}. While such galaxy properties increase the chance of observing LyC emission, they are not sufficient, and the line-of-sight distribution of neutral hydrogen (\hi) gas strongly impacts LyC escape in a given direction. In particular, studies have shown that lines-of-sight with lower \hi\ covering fraction are more favorable to LyC escape \citep{Steidel2018,Gazagnes2018,Gazagnes2020,Saldana-Lopez2022}, and that the global distribution of \hi\ could also play a role in the escape of ionizing emission \citep{LeReste2024,Leclercq2024}.

Several quantities have been investigated as proxies to the LyC emission from galaxies, such as the ratio between the [O{\sc{iii}}]$\lambda 5008$ and [O{\sc{ii}}]$\lambda 3727$ emission lines O$_{32}$, characterizing the ionization state of a galaxy \citep{Jaskot+Oey2013,Chisholm2018,Izotov2018b,Nakajima2020,Flury2022b}. The UV $\beta$ slope is also often used, as it simultaneously traces dust content and the age of stellar populations, both thought to be linked to to the production and escape of ionizing photons \citep{Zackrisson2013,Chisholm2022}.
The properties of the Lyman-alpha (\lya) line have been found to be among one of the best predictors of the escape fraction of ionizing radiation \citep{Verhamme2015,Flury2022b}. In galaxies, \lya\ is commonly emitted following ionization of hydrogen by LyC emission, through recombination. Similarly to LyC radiation, \lya\ emission is strongly impacted by neutral hydrogen and dust. 
Furthermore, \lya\ emission from galaxies has been observed during the EoR \citep[e.g.][]{Witten2024,Tang2024,Witstok2024}, making it a particularly interesting tracer of LyC emission, as it could be used to identify the very galaxies that reionized the Universe. However, most stand-alone galaxy properties, including \lya\ observables, are insufficient to accurately predict the escape fraction of LyC photons \citep{Flury2022a}. Recently, multivariate analysis has shown a tremendous improvement in the prediction of the escape fraction \citep{Jaskot2024a,Choustikov2024}. Such multivariate models are for the first time providing calibrations that can be used to indirectly estimate the LyC escape fraction of galaxies during the EoR, and predict the relative contributions of different galaxy populations to cosmic reionization \citep{Mascia2024,Jaskot2024b}.

Nevertheless, and despite recent improvements on $f_{\rm esc}^{\rm LyC}$ predictions using multivariate models, such predictions still present a large scatter: up to 2 dex for predictions based on single-variables \citep{Flury2022b}, and down to 0.31 dex for the best available multivariate models \citep{Jaskot2024a}. Studies have remained limited by several factors, including the fact that galaxies are objects with complex, three-dimensional structures. 
Few galaxies currently have available sub-kpc scale LyC observations. Notable examples of galaxies with resolved LyC emission are the closest confirmed LyC-emitter Haro 11 \citep{Bergvall2006,Komarova2024}, the lensed $z\sim2.4$ Sunburst Arc \citep{Dahle2016,Rivera-Thorsen2019,Kim2023,Owens2024}, the galaxy with the highest $f_{\rm esc}^{\rm LyC}$ measured, J1316+2614 \citep{Marques-Chaves2024}, and one of the highest redshift (z=3.794) LyC emitter, Ion1 \citep{Ji2025}. In those objects, the LyC emission and escape fraction has been found to vary spatially. In the Sunburst Arc in particular, the majority of escaping LyC emission stems from a single, compact region ($\sim$8 pc) within the galaxy \citep{Mestric2023}. 

High-resolution hydrodynamic simulations provide a complementary view, revealing that LyC and \lya\ escape are both highly anisotropic and time-variable. Several studies have linked bursty feedback to intermittently-higher escape fractions, albeit with a complex dependence on halo mass, star formation history, and ISM structure \citep{Gnedin2008, Wise&Cen2009,Yajima2011,Wise2014,Kimm2014,CenKimm2015,Trebitsch2017,Barrow2020,Ma2020,Ma2021}. Given the large inherent scatter and a desire to understand the global budget for reionization, large-volume simulations (10--100\,cMpc in size) have also been employed to extend the statistical reach of LyC escape studies, e.g. \textsc{FiBy} \citep{Paardekooper2015}, \textsc{sphinx} \citep{Rosdahl2018,Rosdahl2022,Katz2023}, \textsc{thesan} \citep{Yeh2023}, and IllustrisTNG \citep{Kostyuk2023}. Monte Carlo \lya\ radiative transfer post-processing predicts LyC connections between the properties of young ($\lesssim 5$\,Myr) star clusters and galaxy-scale processes, including \hi\ covering fractions, dust and velocity effects, and spatial--spectral features \citep{Smith2019,Smith2022,Blaizot2023,Bhagwat2024}. Beyond this, semi-analytical, numerical experiments, and cloud simulations reveal insights about the porosity in regulating LyC and \lya\ escape \citep{Kakiichi2021,Carr2021,Menon2024}.

Despite this progress, simulations have still not reached consensus on what feedback prescriptions and numerical resolutions are required to properly predict LyC and \lya\ escape, providing further motivation for obtaining constraints from high-resolution observations.
Currently, only globally-integrated galaxy properties of Lyman Continuum Emitters (LCEs) have been measured in a statistically robust manner, even though theoretical and observational frameworks indicate LyC escape to be highly anisotropic.

The Lyman-Alpha and Continuum Origins Survey (LaCOS) has been proposed to address the current lack of information on how resolved galaxy properties influence LyC emission and escape.
LaCOS consists of HST photometry of 42 galaxies part of the extended Low redshift Lyman Continuum Survey \citep[LzLCS+;][]{Flury2022a,Izotov2016a}, obtained in five filters covering the rest-frame UV and optical. The LaCOS images have a $\sim$0.1\arcsec\ PSF, corresponding to a physical scale of $\sim$ $400\,$pc, allowing to relate sub-kpc scale properties to the LyC output of galaxies.
In particular, LaCOS allows for imaging of the \lya\ emission in these galaxies, thought to trace LyC escape. Here, we present the survey, data, and results from photometry. In section \ref{sec:sample}, we detail the sample selection and observations, and present the data products that are being released to the community. In \ref{sec:data-red}, we describe the HST data reduction procedure and present the optical maps of the galaxies in the sample. Section \ref{sec:lya_method} presents the technique employed to obtain \lya\ maps of the galaxies in the sample. In section \ref{sec:corr_fescLyC}, we compare global photometry values to those obtained with COS, and examine whether any trend can be found between sub-kpc scale galaxy properties and the line-of-sight $f_{\rm esc}^{\rm LyC}$ in LaCOS galaxies. We summarize results in section \ref{sec:conclusion}. Additional investigations using LaCOS data include studies of the link between \lya\ halo extent and $f_{\rm esc}^{\rm LyC}$ \citep{Saldana-Lopez2025}, between quantified morphological properties, galaxy merger properties and LyC production and escape (Le Reste et al., in prep.), and will use UV $\beta$ spectral slope maps to investigate the stellar population and dust properties of LaCOS galaxies (Jung et al., in prep.).

This work assumes a standard flat $\Lambda$CDM cosmology with $H_{0}=70\,$km\,s$^{-1}$\,Mpc$^{-1}$, and $\Omega_{m}=0.3$.

\begin{figure*}[t]
    \centering
    \includegraphics[width=\textwidth]{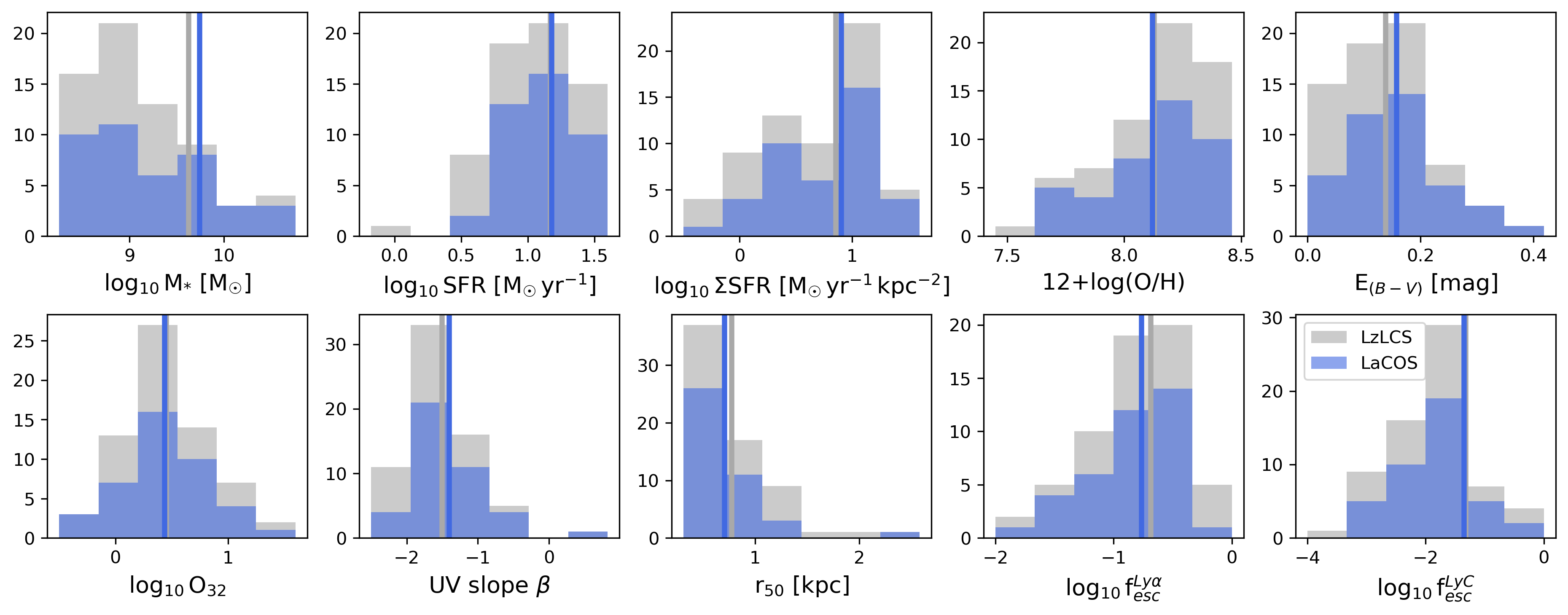}
    \caption{Histogram of properties for the LaCOS (blue) and LzLCS (gray) samples. Vertical bars show the average properties for the two samples (calculated in real space for base 10 log values). The distribution of properties for the LaCOS sample is similar to that of the full LzLCS sample.}
    \label{fig:hist_prop}
\end{figure*}
\section{The Lyman Alpha and Continuum Origins Survey}
\label{sec:sample}
\subsection{Sample Selection}
The LaCOS sample was primarily built from the LzLCS sample, the largest sample of galaxies observed uniformly in LyC at low-redshift \citep[66 galaxies,][]{Flury2022a}, in an effort to understand how the resolved properties of galaxies connects with LyC emission and escape. 
The LzLCS targets were selected from an initial sample of galaxies with available Sloan Digital Sky Survey (SDSS) spectroscopy and Galaxy Evolution Explorer (GALEX) coverage.
Specifically, the LzLCS sample selection targeted nearby galaxies ($z\sim0.3$) with no sign of AGN activity, relatively high ionization states traced via $O_{32}$, high star formation rate surface densities $\Sigma_{\rm SFR}$, or blue UV continua (via their spectral slope index $\beta$). Nevertheless, galaxies were required to fulfill only one of the criteria above to be included in the sample, resulting in a relatively wide range of properties being covered. 
A major goal of the LaCOS survey is to obtain resolved \lya\ maps of objects with LyC observations, to identify whether \lya\ morphology correlates in any way with LyC emission properties. To that effect, the LaCOS sample was constructed from the LzLCS sample, selecting galaxies for which \lya\ imaging could be obtained via the effective narrowband technique. This imaging method, presented in \citet{Hayes2005}, allows the construction of emission line maps from observations with two SBC long-pass filters. One filter samples the emission line and surrounding UV continuum, and the other, only the UV continuum. The emission line map, in this case \lya, can be retrieved by scaling and subtracting one image from the other. This technique makes imaging of \lya\ possible only within a narrow redshift range, for that reason, only LzLCS galaxies with redshift $z<0.32$ are selected. This selection reduced the size of the sample from 66 to 41 galaxies. One additional galaxy from the literature with available archival imaging was added \citep{Izotov2016a}, resulting in a sample of 42 galaxies. 

The final LaCOS sample includes 22 galaxies detected in LyC with a $>2\sigma$ significance \citep[as compared to 35 galaxies in the LzLCS sample, see][]{Flury2022a}, and 20 galaxies with upper limits. The redshift selection thus resulted in an near-even split of LyC emitters and non-emitters, required for statistically robust estimations of the impact of galaxy properties on LyC emission and escape. 

 In Figure \ref{fig:hist_prop}, we present histograms for the properties of LzLCS and LaCOS galaxies, as derived in \citet{Flury2022a}. The distribution of properties and their means are extremely similar for the LzLCS and LaCOS samples. Additionally, we run Kolmogorov--Smirnov tests (using the \texttt{kstest} implementation in the Python package \texttt{scipy}) to verify whether the LaCOS sample is indeed consistent with being drawn from the same distribution as the LzLCS sample. We test each property presented in Figure \ref{fig:hist_prop}, and require a standard $p$-value threshold $p<0.05$ to reject the null hypothesis, which states that the two samples are drawn from the same distribution. Given that $p>0.84$ for all properties considered in Figure \ref{fig:hist_prop}, we cannot reject the null hypothesis. The simple redshift cut operated for the LaCOS sample selection therefore selects galaxies that are representative of the full LzLCS sample.

\subsection{Data}
\label{subsec:HST-data}
The data for the LaCOS program consist of observations in three optical bands with HST WFC3/UVIS filters F850LP, F547M and F438W (with $162$\arcsec$\times162$\arcsec\, field-of-view and $0.04$\arcsec\, pixel scale ), and two UV bands with HST ACS/SBC filters F165LP and F150LP (with $34.6$\arcsec$\times30.5$\arcsec\, field-of-view and $0.032$\arcsec\, pixel scale).
With the redshift cut operated to select the sample, the \lya\ line falls within the HST ACS/SBC F150LP filter, while the F165LP filter samples the UV stellar continuum. The three optical filters were chosen to sample the Balmer break, optical stellar continuum and the broad-band continuum marginally including the \ha\ line for some galaxies. The bulk of data was obtained as part of HST GO program 17069 (PIs Hayes, Scarlata). 

Three of the selected galaxies (J092532, J113304, and J124835) had high-quality archival data available, which were used to complete the sample.
For J124835 and J092532, programs 14131 and 14466 (PI Orlitova) provided observations in all aforementioned instruments and filter combinations, with the exception of observations with filter F850LP, that were executed with instrument ACS/WFC (with $202$\arcsec$\times202$\arcsec\, field of view and 0.05\arcsec\, pixel scale), instead of WFC3/UVIS. Additionally, the ACS/SBC   F150LP filter observations for galaxy J113304 were obtained by HST program 11107 (PI Heckman). 

\begin{figure*}[t]
    \centering
    \includegraphics[width=\textwidth]{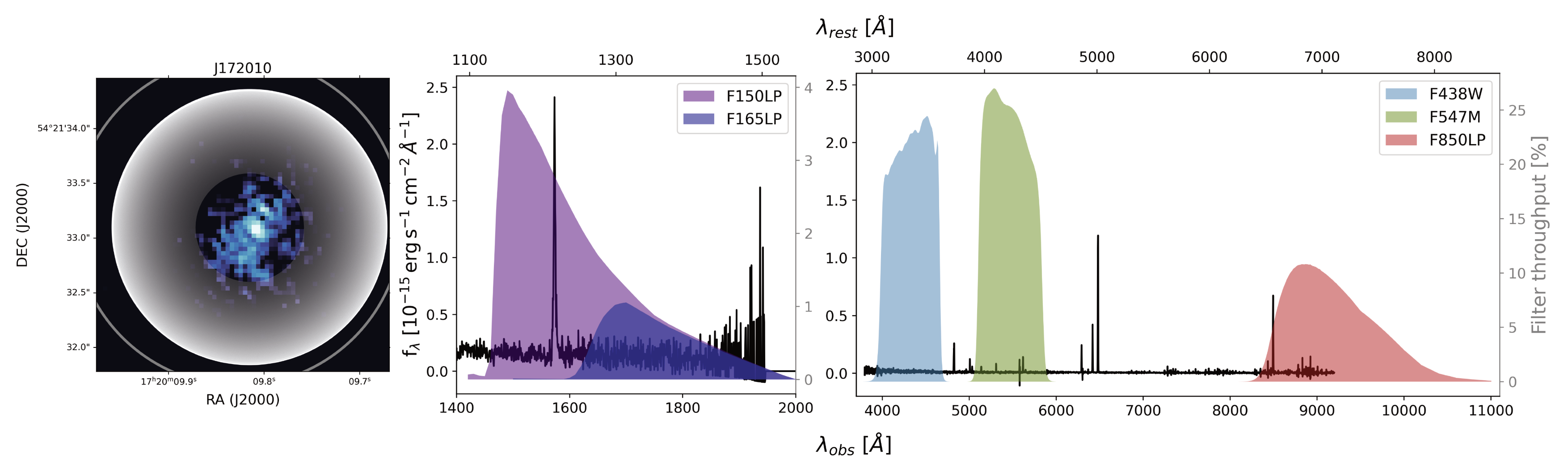}
    \caption{Illustration of LaCOS HST photometry coverage for galaxy J172010 ($z=0.29$). The leftmost panel shows a \lya\ map of the galaxy, with COS (white) and SDSS (gray) extraction apertures overlaid. The panel shows \lya\ flux values between $5 \times 10^{-18}\,$erg$\,$s$^{-1}\,$cm$^{-2}$ (in black) and $1 \times 10^{-16}\,$erg$\,$s$^{-1}\,$cm$^{-2}$ (in white), displayed on a logarithmic scale. The middle panel shows the COS spectrum for wavelengths overlapping with ACS/SBC filter coverage, with filter throughput overlaid. The rightmost panel shows the SDSS spectrum with WFC3/UVIS filter responses overlaid.}
    \label{fig:spec+filt_ex}
\end{figure*}

Exposure times for the different targets and filters, as well as detector temperature for SBC observations are compiled in Appendix Table \ref{tab:hst_obs}. We present an illustration of the HST photometric coverage relative to the COS UV and optical SDSS spectra for galaxy J172010 in Figure \ref{fig:spec+filt_ex}.

We adopt the values described in \cite{Flury2022a} and \cite{Saldana-Lopez2022} for the various galaxy properties presented in this manuscript. These  include the properties shown on in Figure \ref{fig:hist_prop}, and the LyC escape fractions, that are derived from fits to the UV continuum spectral energy distribution. Additionally, and following the convention adopted in LzLCS studies, galaxies are considered to be detected in LyC only if they reach a sufficiently low probability of the measured counts originating from the background $P<0.02275$ (2$\sigma$).

\subsection{Data availability}
\label{sec:data}
The data products for the HST observation of LaCOS galaxies (including archival data) are being released at the Barbara A. Mikulski Archive for Space Telescopes (MAST) as a High Level Science Product, with the following DOI
\dataset[10.17909/ j4qd-ev76]{\doi{10.17909/ j4qd-ev76}}
\footnote{Also accessible via \url{https://archive.stsci.edu/hlsp/lacos/}.}. The released data products include HST images in all individual instrument and filters combination (ACS/SBC F150LP, F165LP, and WFC3/UVIS F438W, F547M and F850LP, or ACS/WFC F850LP). Additionally, it contains \lya\ maps and UV continuum maps for all LaCOS galaxies. Finally, the data release contains a table presenting galaxy properties and photometric measurements, and in particular photometric \lya\ measurements, r$_{50}$ measured in the F165LP filter, star formation rate surface density, and aperture photometry in all filters, extracted in a circular aperture centered on the galaxy coordinates, with radius $ 3\times r_{50}$. In the following, we present the data reduction methods employed to create the imaging data products. 

\section{HST data reduction}
\label{sec:data-red}
Our data reduction approach is inspired from those described in \cite{Runnholm2023} and \cite{Melinder2023}, but has been modified to suit the data available for the LaCOS program. The steps to this approach are described below.

We start the reduction with \texttt{flt} and \texttt{flc} frames downloaded from the MAST archive. These frames have undergone preliminary data reduction with the standard HST reduction pipelines (including bias, initial dark-current subtraction, flat-fielding, and CTE correction) using the latest calibration files.\\

\subsection{Dark current and background subtraction from SBC frames}
\label{sec:dark+bgsub_SBC}
SBC data can be impacted by dark current which manifests as  an elevated, spatially varying background in frames with high detector temperature. While general SBC guidelines have recommended dark current correction to be performed on frames when the detector temperature surpasses 25°C, SBC dark rates have recently been shown to increase before this threshold is reached \citep{GuzmanAvila2024}. Since dark current could mimic the extended \lya\ haloes we are trying to image, removing any dark current contribution from the science frame is imperative. After inspection of the science frames, we set a threshold temperature of $T>22$°C for implementation of the residual dark current subtraction. The method is described in \cite{Runnholm2023}, but in short, we use a collection of dark current files from the same cycles as the observations to simultaneously implement a fit to the sky background and if necessary, dark current pattern in SBC frames using $\chi^2$ minimization. Prior to fitting, the galaxy is masked from the frames using a circular 200 pixel radius mask. From the collection of dark current files, the best-fitting dark frame is identified and scaled to produce a dark current image. We combine this fitted dark current frame with sky background, and subtract those from the data. If the detector temperature is below the threshold temperature ($T<22$°C), only the sky background is subtracted. 

In some instances, the SBC frames still show a gradient background after sky background and/or dark current subtraction. This is a common issue in ACS exposures, linked to the reflection of solar light on the Earth at low limb angles \citep{Biretta2003,Prichard2022}. To correct for this, we model the background as a 2D order 2 polynomial and subtract it from the exposure. The modeling is performed by masking the galaxy using a circular 200 pixel radius mask, and using \texttt{astropy} \texttt{Polynomial2D} model and the Levenberg-Marquardt fitting algorithm. We inspect each background-subtracted frame visually to ensure the background is fully removed before proceeding with further data reduction.

\subsection{Cosmic ray rejection from UVIS and WFC files}
Cosmic rays strongly impact WFC3/UVIS frames. Since we obtained only two UVIS frames per filter, drizzling does not remove all cosmics from the data. To remedy this issue, we run a Python implementation of \texttt{lacosmic} \citep{vanDokkum2001} on the \texttt{flc} frames prior to drizzling to reject cosmic rays. All frames are inspected after cosmic ray removal. Cosmics are successfully removed in most cases, but in a few instances, non-contaminated parts of the galaxy are flagged as cosmic rays. We manually unflag parts of the frame that were wrongly flagged when necessary.

\subsection{Additional background correction for UVIS frames}
Upon further inspection of the frames, we identifiy offsets between the left and right halves of Chip 2 for WFC3/UVIS flc frames. This discontinuity is caused by differences in bias for the amplifiers, and impacts 18\% of the flc frames. Since galaxies are at the center of Chip 2, this discontinuity could affect the fluxes and morphological parameters in the filters impacted. To correct for this, we model and subtract the background separately for the two halves of the chip in the frames impacted. We perform this correction on frames after cosmic ray rejection. Specifically, we use \texttt{photutils} to detect sources and mask them from frames prior to calculating the median background in each half of the chip and subtracting a constant from each frame. A handful of frames present additional structure in their background. In those, we use \texttt{photutils} function \texttt{Background2D} to model a spatially-varying background. All UVIS frames are visually inspected to ensure the background is subtracted correctly.

\subsection{WCS registration and frame alignment}
We use the custom image alignment tool CROCOA\footnote{Publicly available on \url{https://github.com/runnholm/CROCOA}.} developed for the data reduction in \cite{Runnholm2023}. This tool enables fine alignment of frames even in the absence of stars, which is typically the case for the SBC UV frames. Shortly, the code searches for the shifts and rotation in coordinate maximizing the correlation of pixel intensity across different frames. The frames are first aligned in individual filters, and a second run of CROCOA is performed on the pre-aligned frames across the different filters. We visually assess the alignment, adapt the parameters and repeat the process when necessary to ensure all frames are aligned prior to drizzling.

\subsection{Drizzling and additional background subtraction} Once the UVIS and ACS frames are pre-aligned, the frames for individual filters are registered and stacked together to a final pixel scale of $0.04\,$\arcsec\, (the native WFC3 pixel scale) using the HST drizzling software (\texttt{drizzlepac}). For all filters, we produce maps $600\times600$ pix$^2$ ($24$\arcsec$\times24$\arcsec) in size. Since LaCOS galaxies are compact ($r_{50}\sim0.2$\arcsec), these maps fully enclose the emission in all filters for all galaxies. Drizzling also produces weight maps (or inverse variance), that we use to estimate pixel noise.

After drizzling the frames, we subtract any additional background that may be present from the images. We visually inspect the background for all galaxies in each filter individually, and determine the best subtraction scheme, considering several options including no background, a constant background, and 2D polynomials with order from 1 up to 3. Each galaxy is first masked from the frame using a circular mask with 200 pixel radius (8\arcsec). This mask is sufficiently large so that it excludes extended emission from the galaxies, including the \lya\ emission in F150LP images. The masked image, containing only the background, is fit with the function thought to best represent the background. The best-fit to the background is then subtracted from the science frame, and the background-subtracted image is visually inspected to ensure the subtraction was successful. For UVIS filters, the background is in most cases well-described by an order 1 2D polynomial. For SBC frames, most of the galaxies have a background well-approximated by a constant, due to prior background subtraction operated on individual frames (see \ref{sec:dark+bgsub_SBC}). 

\subsection{SBC maps re-weighting} 
Drizzling SBC frames to a larger pixel scale than native leads to noise in drizzled SBC frames being overestimated \citep[see][]{Runnholm2023}. Thus, the SBC frame weights are re-calculated to take into account the correlation of noise across several pixels. To better estimate the noise, we estimated the sky background from the drizzled frames, and then multiplied sky background frames by the factor $R$ \citep{Fruchter2002}, where, given our choice of $0.04\arcsec$ for the final pixel scale $pixscale$:
\begin{equation}
  R  = \frac{1}{1 - \frac{1}{3 r}} \, ,
\end{equation}
with $r=1/pixscale = 1.25$.

\subsection{PSF matching} 
The drizzled frames are matched to a common PSF to enable comparison of the final images in different filters.
Standard PSF matching techniques have problems with matching data sets containing SBC FUV images, due to the extended wings in the SBC PSF.  We use the matching method and PSF models from \cite{Melinder2023}, developed for accurate matching of SBC imaging to optical HST imaging. The PSF models for UV filters are derived for a general case from archival observations of stellar cluster NGC 6681 \citep{Melinder2023}. The images are matched to a common PSF which is constructed from all of the filters to be the broadest PSF at any given radius. In the central pixels, the PSF is dominated by the contribution from the F850LP filter, while in the outer pixels, the PSF contribution is dominated by the UV filters.
After PSF matching, the frames are corrected for the Milky Way extinction using values reported in \cite{Flury2022a}, and the \cite{Fitzpatrick1999} extinction law.

\begin{figure*}[t]
    \centering
    \includegraphics[width=0.99\textwidth]{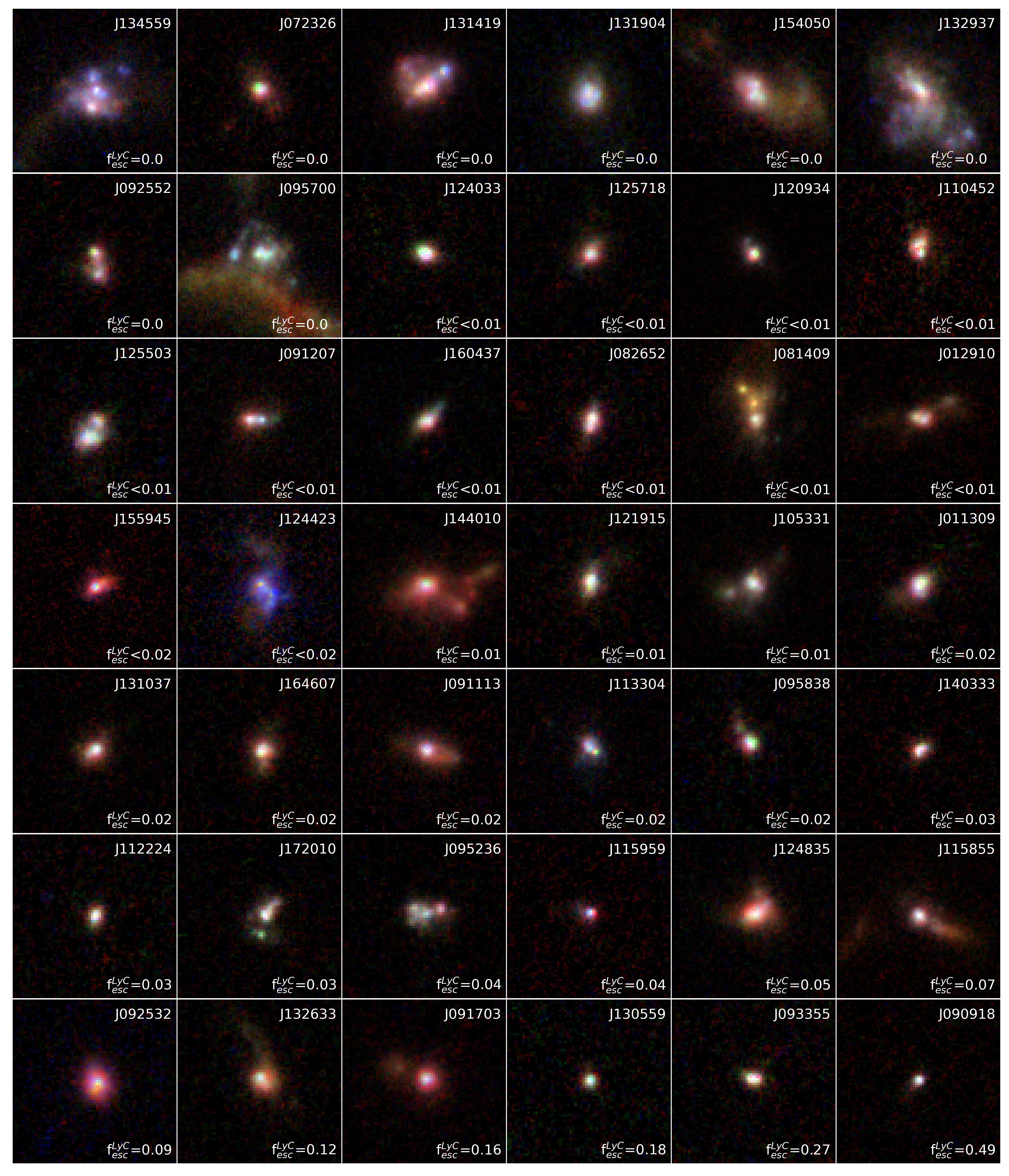}
    \caption{Optical RGB colour-composite images of LaCOS galaxies. The F850LP data is shown in red, F547M in green and F438W in blue. The intensity is normalized to the maximum of the emission in each filter in order to highlight structural differences across different bands, but the bulk of the optical flux is in the F438W filter. The panels are ordered from left to right and top to bottom by increasing $f_{\rm esc}^{\rm LyC}$ (with value shown on the bottom right corner), and all display the galaxies on a scale of $20\,\rm{kpc} \times 20\,\rm{kpc}$, which encompasses the emission in all filters in most cases. Galaxies with emission extending beyond the scale presented in the panels are shown in full in Figure \ref{fig:lacos-large_gals}. LaCOS galaxies show very diverse morphologies, and many exhibit signs of merger interaction.}
    \label{fig:lacos-rgb}.
\end{figure*}

We produce RGB optical colour-composite images for each galaxies, presented in Figure \ref{fig:lacos-rgb}, with F850LP filter in red, F547M in green, and F438W in blue. Each frame is scaled to its own maximum prior to producing the composite image, to highlight structural difference across different bands. The panels display the central $20\times20$ kpc area around the target galaxy, and are sorted by increasing $f_{\rm esc}^{\rm LyC}$.
While most galaxies in the LaCOS sample are physically compact across all filters, a few of them exhibit large-scale features extending beyond the 20$\,$kpc$\times$20$\,$kpc frames presented in Figures \ref{fig:lacos-rgb} and \ref{fig:lacos-lya}, that tend to be more visible in redder filters. In Appendix Figure \ref{fig:lacos-large_gals} we present additional larger-scale optical RGB colour-composite for three galaxies with emission extending beyond the 20 kpc cutouts in Figure \ref{fig:lacos-rgb}.
The resolution reached by HST observations makes it immediately noticeable that LaCOS galaxies have a variety of morphologies, most being irregular, and with several objects having apparent tidal features. We will explore the link between galaxy morphology, merger interaction and LyC emission in a forthcoming manuscript (Le Reste et al. in prep.).

\section{Lyman-alpha maps}
\label{sec:lya_method}
\subsection{Methods}
\label{sec:lya_map_methods}
With 5 photometric bands available, and in the absence of narrow-band imaging covering the H$\alpha$ emission, we cannot employ commonly used photometric SED fitting softwares \citep[such as the \texttt{Lyman Alpha eXtraction Software}, or LaXs,][]{Hayes2009,Melinder2023} to produce \lya\ emission maps.
Nevertheless, we have access to low-resolution spectra of \lya\ and the UV continuum, obtained with HST/COS for all galaxies in the sample, and to imaging in UV filters covering the \lya\ line and the off-line UV continuum. Thus, we can use available imaging and spectra to derive maps reproducing the global \lya\ spectroscopic properties. Note that one galaxy (J124835) has additional photometry, we present a comparison of \lya\ maps obtained with photometric SED fitting using LaXs and the method outlined below in Appendix section \ref{app:LaXs}. Generally, the morphology of the \lya\ and UV continuum maps as well as the UV continuum flux densities derived through the method presented here and via photometric SED fitting with LaXs agrees, however the \lya\ flux (and thus EW) differs significantly, with LaXs values being about half those of the LaCOS and COS measurements. This underestimation of \lya\ flux has been observed previously with LaXs, thus the discrepancy with our measurement is not a source of concern for the quality of data reduction, and we proceed with the method describe below.

\begin{figure*}[t]
    \centering
    \includegraphics[width=0.99\textwidth]{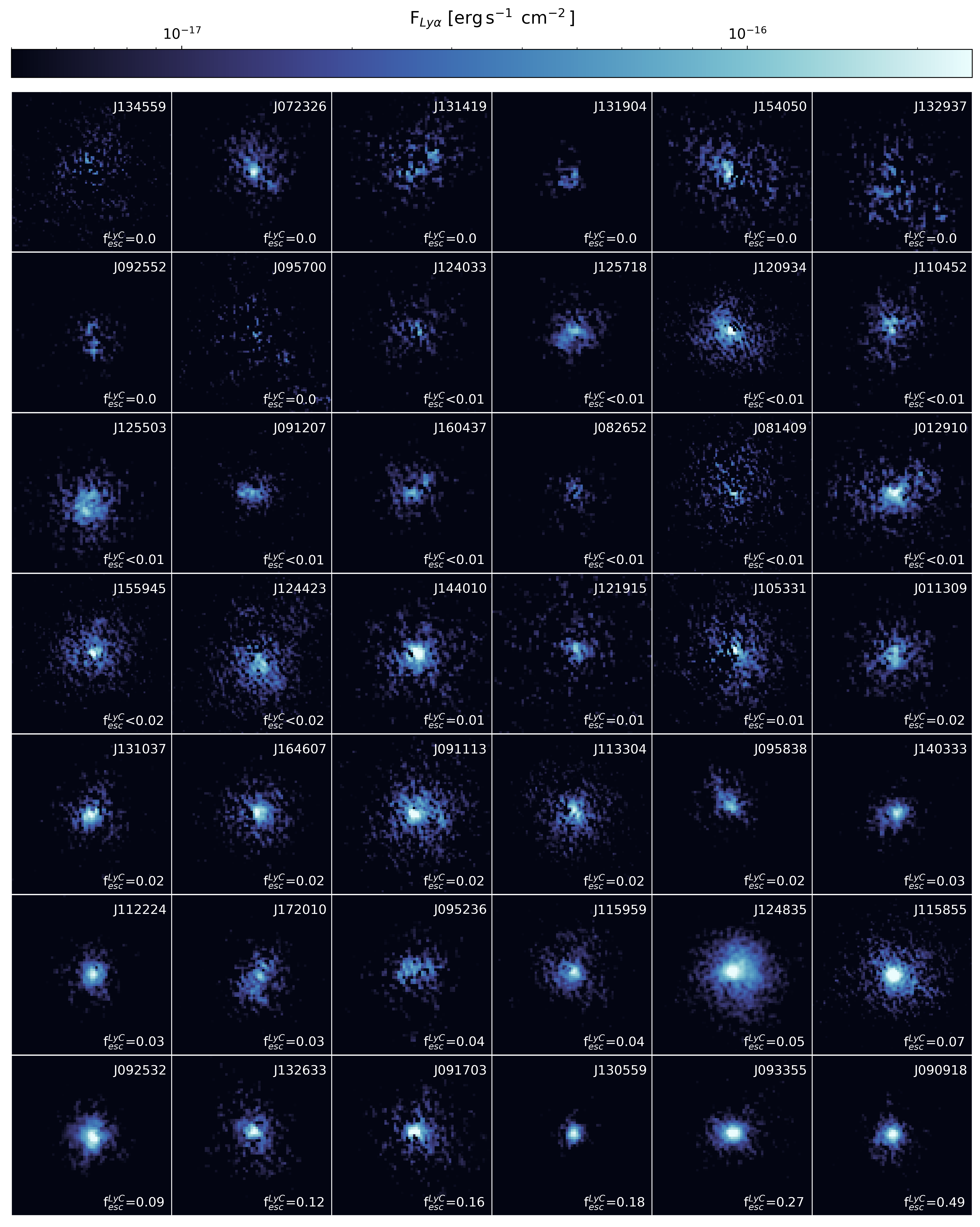}
    \caption{\lya\ maps of the LaCOS galaxies. The panels display a physical scale of $20\,\rm{kpc}\times20\,\rm{kpc}$, and are ordered by increasing LyC escape fraction, displayed in the lower right corner for each panel. For all panels, the \lya\ flux values are shown between $5\times 10^{-18}\,\rm{erg/s/cm}^{2}$ (in black) and $2.5\times 10^{-16}\,\rm{erg/s/cm}^{2}$ (in white), displayed on a logarithmic scale.}
    \label{fig:lacos-lya}
\end{figure*}

\begin{figure*}[t]
    \centering
    \includegraphics[width=0.99\textwidth]{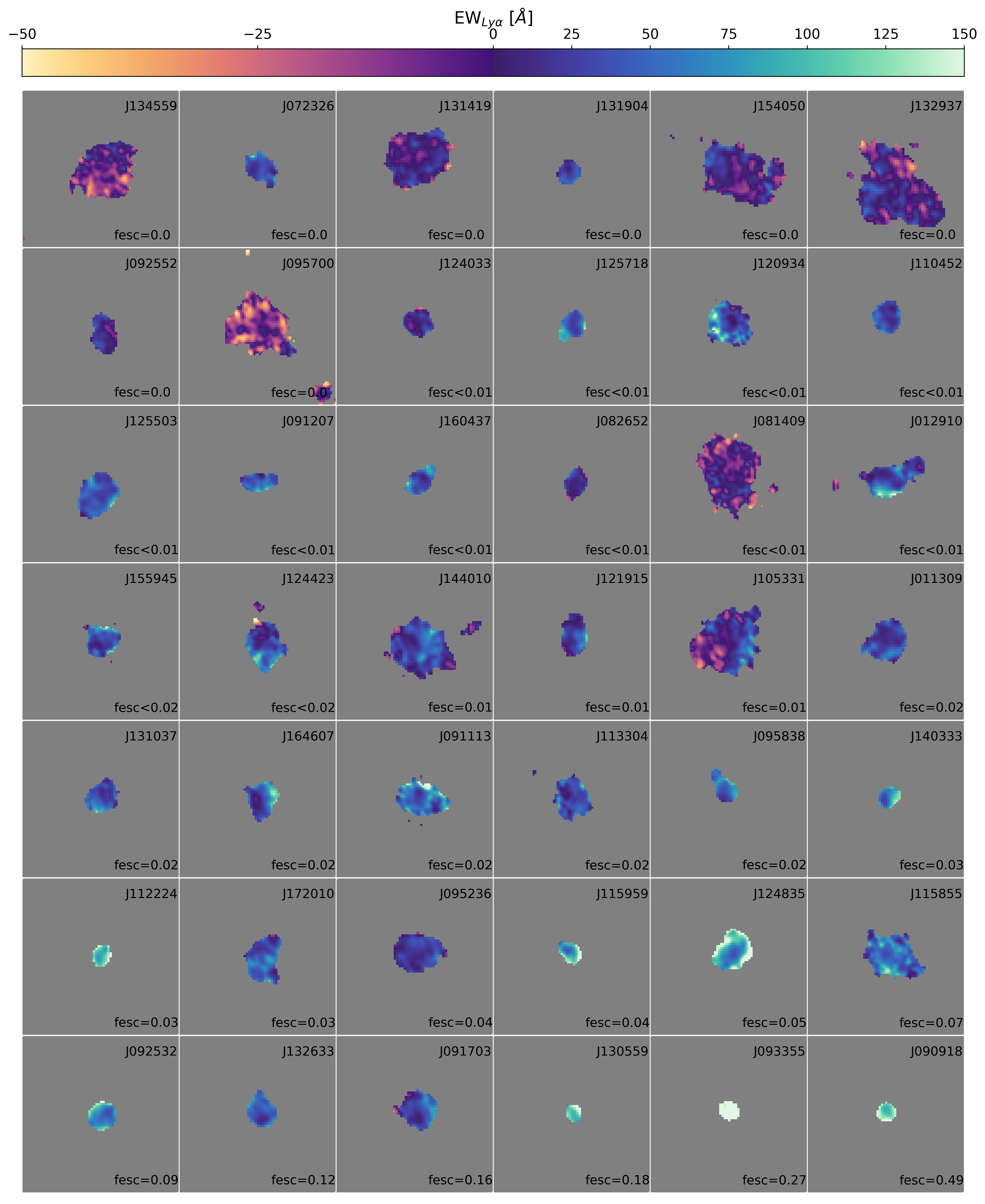}
    \caption{Rest-frame \lya\ EW maps of the LaCOS galaxies. The panels display a physical scale of $20\,\rm{kpc}\times20\,\rm{kpc}$, and are ordered by increasing LyC escape fraction, displayed in the lower right corner for each panel. }
    \label{fig:lacos-ewlya}
\end{figure*}

The LaCOS observations have been designed for the F150LP filter to cover both the \lya\ emission and UV continuum, while the F165LP covers solely the UV continuum. The flux density measured in the F150LP frames can thus be written as a function of the \lya\ line flux $F_{\rm Ly\alpha}$, the UV continuum flux density probed in the F150LP filter $f_{\rm c,F150LP}$, the F165LP filter $f_{\rm{F165LP}}$, and the F150LP bandwidth $\rm{BW}_{\rm{F150LP}}=109.2\,$\AA:
\begin{equation}
 f_{\rm{F150LP}}=\frac{F_{\rm Ly\alpha}}{\rm{BW}_{\rm{F150LP}}} + f_{\rm c,F150LP} \, .
\end{equation}
Here, we introduce the bandpass normalization factor $\alpha$ that links the continuum flux density measured in both filters such that:
\begin{equation}
\label{eq:cont_filters_rel}
f_{\rm c,F150LP}\approx \alpha \cdot f_{\rm{F165LP}} \, .
\end{equation}
The \lya\ images can therefore be obtained once the value of the bandpass normalization factor is known through:
\begin{equation}
\label{eq:lya_phot}
  F_{\rm Ly\alpha} = (f_{\rm{F150LP}} - \alpha \cdot f_{\rm{F165LP}}) \times \rm{BW}_{\rm{F150LP}} \, .
\end{equation}

A simple way of solving for the value of $\alpha$ would be to match the \lya\ photometric flux to the value estimated from the COS spectra. However, to limit possible issues posed by absolute flux calibration errors \citep[theoretically on the order of 2.3\%,][]{Fischer2019}, we instead match the \lya\ image to have the same \lya\ equivalent width (EW) within the COS aperture, as that measured in the COS spectrum. The EW depends on the relative flux between the \lya\ line and UV continuum, and as such, is not impacted by flux calibration uncertainties on COS. However, we note that regardless of performing the match using the spectroscopic EW or $F_{\rm Ly\alpha}$, the resulting \lya\ images could still be impacted by zeropoint uncertainties on ACS/SBC images \citep[1.7\% for F150LP, and 4.5\% for F165LP,][]{Avila2019}, we thus take these uncertainties into account when deriving the error maps. 

The \lya\ EW can be expressed as a function of the \lya\ flux and the continuum flux density value around the \lya\ line $f_{\rm c,Ly\alpha}$ as:
\begin{equation}
\label{eq:ew_def}
 {\rm EW} = \frac{\int f_{\rm Ly\alpha}\, \text{d}\lambda  }{f_{\rm c,Ly\alpha}} = \frac{F_{\rm Ly\alpha}}{f_{\rm c,Ly\alpha}} \, .
\end{equation}
In the photometric case $f_{\rm c,Ly\alpha}=f_{\rm c,F150LP}$, so we combine Eqs.~(\ref{eq:cont_filters_rel}) and (\ref{eq:lya_phot}) to write:
\begin{equation}
\label{eq:ew_phot}
    {\rm EW}_{\rm phot} = \left(\frac{1}{\alpha}\cdot\frac{f_{\rm F150LP}}{f_{\rm F165LP}}-1\right)\cdot  \rm{BW}_{\rm{F150LP}} \, .
\end{equation}
Since we have access to the value of the rest-frame \lya\ equivalent width EW$_{\rm COS}$ in the COS aperture \citep{Flury2022a}, we can solve for $\alpha$ by extracting photometry in the same aperture as that used to extract the COS spectra. To do so, we use a vignetted circular aperture with 2.5\arcsec\ diameter (see the example COS extraction aperture shown on the left panel of Figure \ref{fig:spec+filt_ex}, and the discussion in Appendix \ref{app:COSap}). 
We center the aperture upon the brightest pixel in the F165LP image. We sum the photometric \lya\ flux and the continuum flux density within the aperture to obtain a photometric equivalent width matched to that observed with COS\footnote{The spectroscopic \lya\ flux and EW we use here for galaxy J092532 have been updated from the values presented in \cite{Flury2022a} to $F_{\rm Ly\alpha}=2.20\pm 0.09 \times 10^{-14}\,\rm{erg}\,\rm{s}^{-1}\,\rm{cm}^{-2}$ and $EW=74.6 \pm5.8\,$\AA\ as an internal dust correction had been applied in \cite{Izotov2016a} where the values were reported, which had not been taken into account previously.}, ${\rm EW}_{\rm COS,obs} = (1+z) \cdot {\rm EW}_{\rm COS}$, and derive the bandpass normalization factor $\alpha$ as:

\begin{equation}
\alpha = \frac{f_{\rm F150LP,ap}}{f_{\rm F165LP,ap}} \cdot \frac{ \rm{BW}_{\rm{F150LP}}}{{\rm EW}_{\rm COS,obs}+ \rm{BW}_{\rm{F150LP}}} \, .
\end{equation}
We use the $\alpha$ values (0.63--1.1) derived by matching the photometric and spectroscopic EW to derive \lya\ maps using Eq.~(\ref{eq:lya_phot}), UV continuum maps using Eq.~(\ref{eq:cont_filters_rel}), and EW maps using Eq.~(\ref{eq:ew_def}). Error maps are derived via error propagation taking into account the errors on $\alpha$, uncertainty on pixel flux and zero-point uncertainties on photometry.

\subsection{Ly\texorpdfstring{$\alpha$}{a} maps}
We present the \lya\ maps obtained following the approach described in \ref{sec:lya_map_methods} in Figure \ref{fig:lacos-lya}. Additionally, we show EW maps in Figure \ref{fig:lacos-ewlya}, produced by smoothing the \lya\ and UV continuum maps with Gaussian kernel with 2.5 pixel full-width half-maximum (FWHM), and masking regions with $f_{\rm \lambda,UV} < 1.5 \times 10^{-19}\,$erg$\,$s$^{-1}\,$cm$^{-2}\,$\AA$^{-1}$, prior to diving the \lya\ by the continuum maps. Similarly to Figure \ref{fig:lacos-rgb}, all panels have a physical size of $20\,\rm{kpc}\times20\,\rm{kpc}$, and are sorted from top to bottom and left to right by LyC escape fraction. Additionally, all maps are shown on the same logarithmic scales. The galaxies with highest $f_{\rm esc}^{\rm LyC}$ tend to have brighter and more compact \lya\ emission. Furthermore, galaxies with high $f_{\rm esc}^{\rm LyC}$ tend to have larger EW ($\geq50$\AA), while non-LyC emitting galaxies often show low and negative EW values, indicative of \lya\ absorption, linked to the presence of neutral gas or dust on the line-of-sight. Nevertheless, the \lya\ emission intensity, morphology and extent  varies strongly across the sample. Detailed analysis of the \lya\ halo morphology and the relation with LyC output is presented in \cite{Saldana-Lopez2025}. In the following, we explore the link between global \lya\ photometry measured in the SBC maps and the LyC escape fraction.

In Table \ref{tab:LaCOS-sample} we present galaxy properties, including measurements from \lya\ photometry for the LaCOS sample. The redshift and $f_{\rm esc}^{\rm LyC}$ are obtained from \citet{Flury2022a}. The \lya\ properties, UV continuum flux density $f_{\lambda,\rm{UV}}$ and $r_{50}$ values are derived through SBC photometry.

\begin{table*}[h]
    \centering
    \small
        \caption{LaCOS galaxy properties and photometric measurements.}
    \begin{tabular}{cccccccccc}
    \hline
ID & z\tablenotemark{a} & $f_{\rm esc}^{\rm LyC}$\tablenotemark{a} & $L_{\rm Ly\alpha}$ & EW$_{\rm Ly\alpha}$ & $f_{\rm esc}^{\rm Ly\alpha}$ & $f_{\lambda,\rm{UV}}$ & r$_{50}$ & $\Sigma_{\rm{\rm SFR}}$ \\ 
 &  &  & $10^{42}\,$erg$\,$s$^{-1}$ & \AA & & $10^{-16}\,$erg$\,$s$^{-1}\,$cm$^{-2}\,$\AA$^{-1}$ & kpc & M$_{\odot}\,$yr$^{-1}\,$kpc$^{-2}$ \\
\hline
J011309  &  0.31  &  0.02 $_{- 0.01 }^{+ 0.02 }$  &  4.84 $\pm$ 0.33  &  47.8 $\pm$ 3.6  &  0.65 $\pm$ 0.12  &  2.56 $\pm$ 0.09  &  0.8 $\pm$ 0.09  &  1.06 $\pm$ 0.37  \\ 
J012910  &  0.28  &   $<$ 0.007  &  5.4 $\pm$ 0.44  &  49.7 $\pm$ 4.6  &  0.23 $\pm$ 0.02  &  3.45 $\pm$ 0.15  &  0.9 $\pm$ 0.08  &  2.4 $\pm$ 0.48  \\ 
J072326  &  0.3  &   $<$ 0.004  &  3.16 $\pm$ 0.18  &  55.2 $\pm$ 3.7  &  0.23 $\pm$ 0.02  &  1.57 $\pm$ 0.05  &  0.7 $\pm$ 0.09  &  2.69 $\pm$ 0.75  \\ 
J081409  &  0.23  &   $<$ 0.008  &  0.26 $\pm$ 0.22  &  1.3 $\pm$ 1.1  &  0.01 $\pm$ 0.01  &  10.5 $\pm$ 0.12  &  1.8 $\pm$ 0.07  &  0.76 $\pm$ 0.08  \\ 
J082652  &  0.3  &   $<$ 0.009  &  1.45 $\pm$ 0.44  &  67.1 $\pm$ 24.8  &  0.14 $\pm$ 0.04  &  0.59 $\pm$ 0.13  &  0.8 $\pm$ 0.09  &  1.64 $\pm$ 0.45  \\ 
J090918  &  0.28  &  0.49 $_{- 0.23 }^{+ 0.42 }$  &  3.47 $\pm$ 0.34  &  202.1 $\pm$ 47.8  &  0.22 $\pm$ 0.03  &  0.54 $\pm$ 0.12  &  0.5 $\pm$ 0.09  &  6.15 $\pm$ 2.15  \\ 
J091113  &  0.26  &  0.02 $_{- 0.01 }^{+ 0.02 }$  &  6.18 $\pm$ 0.25  &  61.8 $\pm$ 2.9  &  0.14 $\pm$ 0.01  &  3.74 $\pm$ 0.09  &  0.6 $\pm$ 0.08  &  9.58 $\pm$ 2.43  \\ 
J091207  &  0.25  &   $<$ 0.008  &  1.71 $\pm$ 0.2  &  55.2 $\pm$ 7.4  &  0.08 $\pm$ 0.01  &  1.34 $\pm$ 0.09  &  1.4 $\pm$ 0.08  &  1.05 $\pm$ 0.14  \\ 
J091703  &  0.3  &  0.16 $_{- 0.06 }^{+ 0.07 }$  &  5.91 $\pm$ 0.42  &  44.9 $\pm$ 3.5  &  0.18 $\pm$ 0.02  &  3.49 $\pm$ 0.12  &  0.5 $\pm$ 0.09  &  10.92 $\pm$ 3.69  \\ 
J092532  &  0.3  &  0.09 $_{- 0.03 }^{+ 0.02 }$  &  5.08 $\pm$ 0.16  &  80.7 $\pm$ 3.0  &  0.14 $\pm$ 0.01  &  1.66 $\pm$ 0.04  &  0.6 $\pm$ 0.09  &  9.45 $\pm$ 2.73  \\ 
J092552  &  0.31  &   $<$ 0.004  &  1.29 $\pm$ 0.19  &  30.1 $\pm$ 4.6  &  0.15 $\pm$ 0.02  &  1.01 $\pm$ 0.05  &  1.2 $\pm$ 0.09  &  0.65 $\pm$ 0.16  \\ 
J093355  &  0.29  &  0.27 $_{- 0.11 }^{+ 0.11 }$  &  5.26 $\pm$ 0.14  &  286.2 $\pm$ 19.1  &  0.4 $\pm$ 0.04  &  0.53 $\pm$ 0.03  &  0.7 $\pm$ 0.09  &  2.77 $\pm$ 0.78  \\ 
J095236  &  0.32  &  0.04 $_{- 0.01 }^{+ 0.02 }$  &  3.05 $\pm$ 0.44  &  35.6 $\pm$ 5.5  &  0.28 $\pm$ 0.05  &  1.96 $\pm$ 0.11  &  1.1 $\pm$ 0.09  &  0.89 $\pm$ 0.2  \\ 
J095700  &  0.24  &   $<$ 0.001  &  -0.99 $\pm$ 0.13  &  -10.9 $\pm$ 1.4  &  -0.02 $\pm$ 0.0  &  4.02 $\pm$ 0.06  &  8.0 $\pm$ 0.23  &  0.07 $\pm$ 0.0  \\ 
J095838  &  0.3  &  0.02 $_{- 0.01 }^{+ 0.03 }$  &  2.21 $\pm$ 0.25  &  89.3 $\pm$ 13.7  &  0.09 $\pm$ 0.01  &  0.65 $\pm$ 0.07  &  1.4 $\pm$ 0.09  &  1.2 $\pm$ 0.17  \\ 
J105331  &  0.25  &  0.01 $_{- 0.0 }^{+ 0.01 }$  &  4.77 $\pm$ 0.48  &  17.5 $\pm$ 1.8  &  0.1 $\pm$ 0.01  &  11.2 $\pm$ 0.21  &  0.8 $\pm$ 0.08  &  7.03 $\pm$ 1.43  \\ 
J110452  &  0.28  &   $<$ 0.011  &  3.44 $\pm$ 0.16  &  63.9 $\pm$ 3.6  &  0.49 $\pm$ 0.05  &  1.71 $\pm$ 0.05  &  0.9 $\pm$ 0.08  &  0.86 $\pm$ 0.25  \\ 
J112224  &  0.3  &  0.03 $_{- 0.02 }^{+ 0.06 }$  &  3.33 $\pm$ 0.4  &  232.8 $\pm$ 69.6  &  0.28 $\pm$ 0.04  &  0.37 $\pm$ 0.1  &  0.6 $\pm$ 0.09  &  3.23 $\pm$ 1.01  \\ 
J113304  &  0.24  &  0.02 $_{- 0.01 }^{+ 0.02 }$  &  4.95 $\pm$ 0.32  &  53.3 $\pm$ 3.9  &  0.43 $\pm$ 0.04  &  4.27 $\pm$ 0.15  &  1.0 $\pm$ 0.08  &  1.21 $\pm$ 0.25  \\ 
J115855  &  0.24  &  0.07 $_{- 0.02 }^{+ 0.03 }$  &  7.91 $\pm$ 0.3  &  50.7 $\pm$ 2.1  &  0.22 $\pm$ 0.01  &  7.03 $\pm$ 0.14  &  0.7 $\pm$ 0.08  &  7.12 $\pm$ 1.62  \\ 
J115959  &  0.27  &  0.04 $_{- 0.02 }^{+ 0.07 }$  &  3.81 $\pm$ 0.2  &  168.9 $\pm$ 17.9  &  0.35 $\pm$ 0.03  &  0.8 $\pm$ 0.07  &  0.8 $\pm$ 0.08  &  1.66 $\pm$ 0.41  \\ 
J120934  &  0.22  &   $<$ 0.013  &  5.84 $\pm$ 0.27  &  54.8 $\pm$ 2.9  &  0.19 $\pm$ 0.01  &  6.19 $\pm$ 0.16  &  0.6 $\pm$ 0.07  &  9.41 $\pm$ 2.41  \\ 
J121915  &  0.3  &  0.01 $_{- 0.0 }^{+ 0.02 }$  &  1.99 $\pm$ 0.4  &  44.5 $\pm$ 10.0  &  0.21 $\pm$ 0.04  &  1.16 $\pm$ 0.11  &  1.0 $\pm$ 0.09  &  0.99 $\pm$ 0.25  \\ 
J124033  &  0.28  &   $<$ 0.011  &  1.87 $\pm$ 0.2  &  37.8 $\pm$ 4.4  &  0.11 $\pm$ 0.01  &  1.53 $\pm$ 0.07  &  0.7 $\pm$ 0.09  &  4.02 $\pm$ 1.07  \\ 
J124423  &  0.24  &   $<$ 0.015  &  6.15 $\pm$ 0.35  &  71.1 $\pm$ 5.0  &  0.11 $\pm$ 0.01  &  4.06 $\pm$ 0.17  &  1.4 $\pm$ 0.08  &  3.03 $\pm$ 0.35  \\ 
J124835  &  0.26  &  0.05 $_{- 0.03 }^{+ 0.04 }$  &  10.96 $\pm$ 0.14  &  113.9 $\pm$ 2.0  &  0.4 $\pm$ 0.02  &  3.56 $\pm$ 0.04  &  0.7 $\pm$ 0.08  &  5.18 $\pm$ 1.19  \\ 
J125503  &  0.31  &   $<$ 0.009  &  5.18 $\pm$ 0.32  &  63.9 $\pm$ 4.8  &  0.31 $\pm$ 0.03  &  1.96 $\pm$ 0.08  &  1.6 $\pm$ 0.09  &  0.64 $\pm$ 0.09  \\ 
J125718  &  0.31  &   $<$ 0.014  &  2.7 $\pm$ 0.18  &  65.2 $\pm$ 5.2  &  0.12 $\pm$ 0.01  &  0.99 $\pm$ 0.04  &  0.7 $\pm$ 0.09  &  4.11 $\pm$ 1.08  \\ 
J130559  &  0.32  &  0.18 $_{- 0.06 }^{+ 0.08 }$  &  1.64 $\pm$ 0.3  &  182.9 $\pm$ 69.0  &  0.14 $\pm$ 0.03  &  0.21 $\pm$ 0.07  &  0.6 $\pm$ 0.09  &  3.8 $\pm$ 1.38  \\ 
J131037  &  0.28  &  0.02 $_{- 0.01 }^{+ 0.02 }$  &  3.55 $\pm$ 0.24  &  50.4 $\pm$ 3.9  &  0.18 $\pm$ 0.02  &  2.18 $\pm$ 0.08  &  0.6 $\pm$ 0.09  &  5.49 $\pm$ 1.64  \\ 
J131419  &  0.3  &   $<$ 0.001  &  2.31 $\pm$ 0.49  &  13.0 $\pm$ 2.8  &  0.06 $\pm$ 0.01  &  4.9 $\pm$ 0.14  &  2.0 $\pm$ 0.09  &  0.87 $\pm$ 0.09  \\ 
J131904  &  0.32  &   $<$ 0.002  &  0.74 $\pm$ 0.16  &  30.3 $\pm$ 6.8  &  0.03 $\pm$ 0.01  &  0.57 $\pm$ 0.04  &  1.2 $\pm$ 0.09  &  1.97 $\pm$ 0.32  \\ 
J132633  &  0.32  &  0.12 $_{- 0.08 }^{+ 0.14 }$  &  4.22 $\pm$ 0.3  &  60.6 $\pm$ 5.2  &  0.13 $\pm$ 0.01  &  1.61 $\pm$ 0.07  &  0.7 $\pm$ 0.09  &  6.2 $\pm$ 1.58  \\ 
J132937  &  0.31  &   $<$ 0.001  &  1.53 $\pm$ 0.53  &  7.5 $\pm$ 2.6  &  0.06 $\pm$ 0.02  &  5.01 $\pm$ 0.14  &  3.0 $\pm$ 0.09  &  0.31 $\pm$ 0.03  \\ 
J134559  &  0.24  &   $<$ 0.002  &  1.45 $\pm$ 0.4  &  8.8 $\pm$ 2.4  &  0.05 $\pm$ 0.01  &  7.94 $\pm$ 0.2  &  1.7 $\pm$ 0.08  &  0.89 $\pm$ 0.1  \\ 
J140333  &  0.28  &  0.03 $_{- 0.01 }^{+ 0.02 }$  &  2.28 $\pm$ 0.18  &  86.7 $\pm$ 9.0  &  0.12 $\pm$ 0.01  &  0.82 $\pm$ 0.06  &  0.7 $\pm$ 0.09  &  4.2 $\pm$ 1.11  \\ 
J144010  &  0.3  &  0.01 $_{- 0.0 }^{+ 0.0 }$  &  8.36 $\pm$ 0.43  &  39.5 $\pm$ 2.2  &  0.14 $\pm$ 0.01  &  5.61 $\pm$ 0.12  &  0.9 $\pm$ 0.09  &  7.13 $\pm$ 1.44  \\ 
J154050  &  0.29  &   $<$ 0.001  &  3.78 $\pm$ 0.41  &  15.2 $\pm$ 1.7  &  0.09 $\pm$ 0.01  &  6.97 $\pm$ 0.12  &  1.1 $\pm$ 0.09  &  3.04 $\pm$ 0.48  \\ 
J155945  &  0.23  &   $<$ 0.025  &  4.39 $\pm$ 0.18  &  56.2 $\pm$ 2.7  &  0.27 $\pm$ 0.02  &  4.18 $\pm$ 0.1  &  0.7 $\pm$ 0.07  &  3.76 $\pm$ 0.92  \\ 
J160437  &  0.31  &   $<$ 0.007  &  2.56 $\pm$ 0.37  &  106.5 $\pm$ 23.0  &  0.07 $\pm$ 0.01  &  0.58 $\pm$ 0.09  &  0.9 $\pm$ 0.09  &  4.13 $\pm$ 0.85  \\ 
J164607  &  0.29  &  0.02 $_{- 0.01 }^{+ 0.01 }$  &  4.82 $\pm$ 0.45  &  65.0 $\pm$ 7.5  &  0.17 $\pm$ 0.02  &  2.14 $\pm$ 0.14  &  0.7 $\pm$ 0.09  &  6.25 $\pm$ 1.6  \\ 
J172010  &  0.29  &  0.03 $_{- 0.01 }^{+ 0.03 }$  &  3.11 $\pm$ 0.34  &  65.0 $\pm$ 8.6  &  0.22 $\pm$ 0.03  &  1.35 $\pm$ 0.1  &  1.4 $\pm$ 0.09  &  0.71 $\pm$ 0.12  \\ 
        \hline
    \end{tabular}
    \tablenotetext{a}{Values taken from \citet{Flury2022a}.}
    \label{tab:LaCOS-sample}
\end{table*}

\subsection{Comparison to spectroscopic Ly\texorpdfstring{$\alpha$}{a} measurements}
\begin{figure*}[ht]
    \centering
    \includegraphics[width=\textwidth]{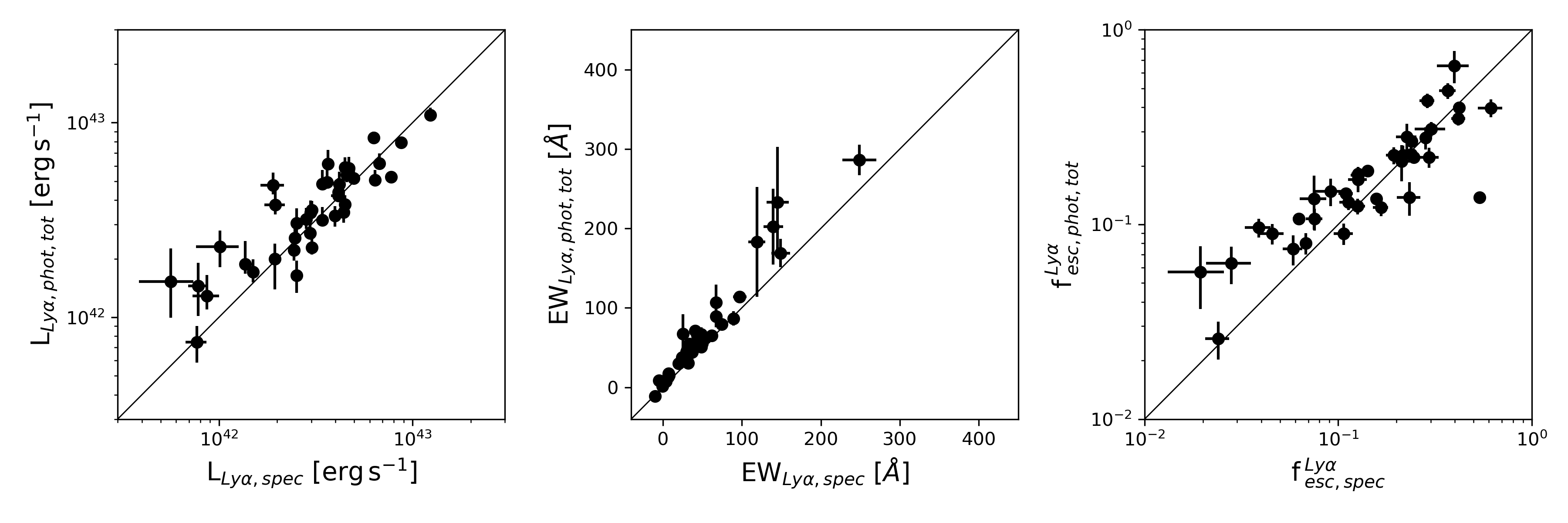}
    \caption{Comparison between \lya\ observables obtained via COS spectroscopy and global photometry. From left to right the panels show comparisons for the \lya\ luminosity ($L_{\rm Ly\alpha}$), rest-frame equivalent width (EW$_{\rm Ly\alpha}$), and escape fraction ($f_{\rm esc}^{\rm Ly\alpha}$). The solid line show the 1:1 relation. The global values from photometry are in broad agreement with the spectral values, indicating the galaxies are compact.}
    \label{fig:lya_specvphot}
\end{figure*}

In Figure \ref{fig:lya_specvphot}, we compare the \lya\ measurements obtained from the  COS spectra to the global \lya\ measurements from SBC photometry, specifically the \lya\ luminosity, EW, and escape fraction $f_{\rm esc}^{\rm Ly\alpha}$. The Ly$\alpha$ flux is measured through curve-of-growth integration within circular apertures, each spaced by 5 pixels (0.2\arcsec). Using this approach, we measure the flux in increasingly larger apertures, and set the aperture for integration when the flux reaches a plateau (with less than 0.1\% fractional variation in flux), or to the radius where the \lya\ flux reaches a maximum. This circular aperture is used to measure the \lya\ flux, EW and $f_{\rm esc}^{\rm Ly\alpha}$, and has an average radius of 173 pixels (6.9\arcsec). We calculate the rest-frame EW from the observed values, such that ${\rm EW} = {\rm EW}_{\rm obs}/(1+z)$. The \lya\ escape fraction $f_{\rm esc}^{\rm Ly\alpha}$ is obtained by dividing the observed photometric \lya\ flux by the intrinsic \lya\ flux. The latter value is estimated using dust-corrected \hb\ flux derived from SDSS spectra, such that
\begin{equation}
f_{\rm esc}^{\rm Ly\alpha}=\frac{F_{\rm Ly\alpha}}{\eta(n_e,T_e)\times F_{\rm H\beta, dustcorr}} \, .
\end{equation}
The values for the ratio between intrinsic \lya\ and \hb\ fluxes $\eta(n_e,T_e)$ are calculated using the python package \texttt{Pyneb} \citep{Luridiana2015}. This calculation uses the dust-corrected \hb\ flux, $n_e$ and $T_e$ measurements presented in \citet{Flury2022a}.

Generally, the \lya\ observables obtained via global photometry are slightly larger than the values obtained in COS spectroscopy. A few galaxies have \lya\ luminosities and escape fractions lower with global photometry than the spectroscopic value. After inspection of the spectra and images, we deduce this might be caused by a combination of UV continuum placement in spectroscopy potentially impacting the EW, and to uncertainties on the zero-point offsets in imaging. If the spectroscopic EW is overestimated, and/or if the zero point offset in the imaging leads to under-estimating the UV continuum flux, the \lya\ emission within the COS aperture in photometry would be fainter than what is measured with spectroscopy. In a compact galaxy, this would produce a global photometric \lya\ flux that is lower than what is measured in COS spectra. However, due to the general agreement between the \lya\ luminosity values, we determine this is not a significant issue. Additionally, the relative agreement between the photometric and spectroscopic values indicates that the \lya\ emission in LaCOS galaxies is compact: in most cases, the bulk of the \lya\ emission is observed within the 2.5\arcsec\ COS aperture.

\section{Correlations with LyC escape}
\label{sec:corr_fescLyC}
Here we compare the global galaxy properties measured with HST photometry to $f_{\rm esc}^{\rm LyC}$ measured within the COS aperture. Specifically, we look at properties previously investigated via COS spectroscopy and through the COS acquisition images \citep{Flury2022b}. We use the Kendall $\tau$ coefficient and associated $p$-value to assess the degree of correlation between variables, using the scheme developed in \citet{Isobe1986} that allows for the inclusion of censored data\footnote{We use the code developed in \citet{Herenz2024} and adapted from \citep{Flury2022b}, publicly available on \url{https://github.com/Knusper/kendall}.}. In Table \ref{tab:correlations_COSvSBC} we show the Kendall $\tau$ and $p$-value between $f_{\rm esc}^{\rm LyC}$ and different galaxy properties obtained for the LaCOS sample with either COS or SBC.  Following the convention in LzLCS studies, we deem a correlation ($\tau>0$) or anti-correlation ($\tau<0$) significant when $p<0.00135$ \citep{Flury2022b}, corresponding to a $3\sigma$ confidence that the null hypothesis is rejected. Additionally, we describe correlations as tentative when $0.00135<p<0.05$.

\subsection{Global Ly\texorpdfstring{$\alpha$}{a} photometry and LyC escape }
\label{sec:LyaphotVSfesc}
\begin{figure*}[ht]
    \centering
    \includegraphics[width=\textwidth]{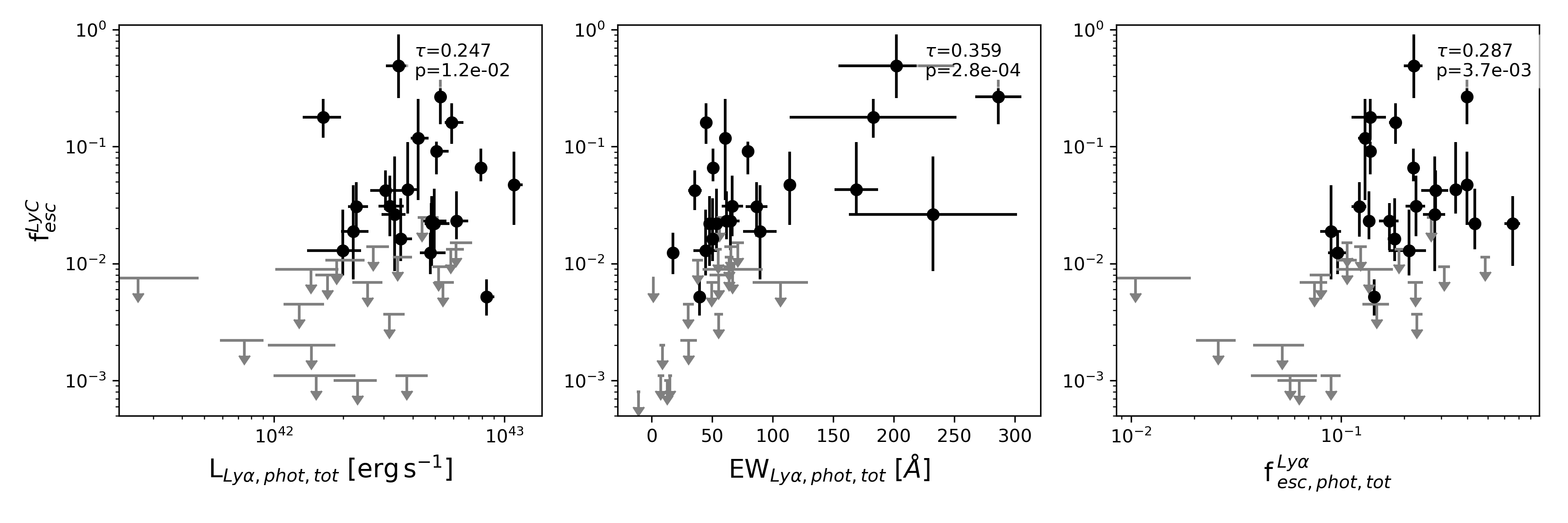}
    \caption{Global photometric \lya\ observables vs $f_{\rm esc}^{\rm LyC}$. From left to right, the $x$-axis shows the \lya\ luminosity, EW and \lya\ escape fraction. Black filled circles show galaxies confidently detected in LyC, while gray arrows show galaxies with LyC escape fraction upper limits. Kendall $\tau$ and $p$-values assessing the degree of correlation between $x$ and $y$ axes variables are shown on the top right corner of each panel. 
    The escape fraction shows the same trends with photometric \lya\ observables as those identified with the spectroscopic values \citep{Flury2022b}, but the correlation with \lya\ luminosity and escape fraction are weaker, while $f_{\rm esc}^{\rm LyC}$ shows a slightly larger degree of correlation with the photometric \lya\ EW.}
    \label{fig:lya_photvfesc}
\end{figure*}

Here we directly show how slight changes in the \lya\ observables when obtained via photometry impact correlations with the LyC escape fraction. In Figure \ref{fig:lya_photvfesc} we show scatter plots comparing \lya\ observables (L$_{\rm Ly\alpha}$, EW$_{\rm Ly\alpha}$ and $f_{\rm esc}^{\rm Ly\alpha}$) to $f_{\rm esc}^{\rm LyC}$. 
We find a significant correlation between $f_{\rm esc}^{\rm LyC}$ and photometric EW$_{\rm Ly\alpha}$, and tentative correlations with $L_{\rm Ly\alpha}$ and $f_{\rm esc}^{\rm Ly\alpha}$. As compared to spectroscopic values within the COS aperture, the global \lya\ luminosity and escape fractions show a decreased degree of correlation with the LyC escape fraction, as probed by Kendall $\tau$ coefficient and associated $p$-values. This is sensible, as the \lya\ photometric values generally probe emission on larger scales than those measured with the COS aperture, not necessarily representing the \lya\ properties of the regions preferentially leaking LyC in the galaxy. The global photometric EW, however, shows a slightly larger degree of correlation with the escape fraction ($\tau=0.41,p=4.1e-5$) as compared to the spectroscopic values ($\tau=0.38,p=9.9e-5$). Nevertheless, the change is relatively small and can be accounted for by the uncertainty on photometric values and general uncertainties on $\tau$ for spectroscopic values \citep[$\sim 0.05$][]{Flury2022b}.
In the following, we look at the impact of sub-kpc scale \lya\ properties on LyC escape. 

\subsection{Resolved Ly\texorpdfstring{$\alpha$}{a} properties and LyC escape }
Few studies have been able to observe the impact of resolved galaxy properties onto $f_{\rm esc}^{\rm LyC}$. The first study looking at the impact of resolved \lya\ properties on LyC escape fractions was conducted on the Sunburst Arc, a lensed galaxy at $z\sim2.4$ \citep{Kim2023}. In this object, LyC emission originates from a single compact star-forming region, that appears to be a star cluster  $r_{\rm eff}\sim8$pc in size \citep{Mestric2023,Rivera-Thorsen2024}. The gravitational lens has produced 12 distinct images of this region, all sampling different lines of sight to the LyC-emitting cluster. The LyC-emitting region has extreme \lya\ properties compared to non-emitting regions in the rest of the galaxy, and compared to the global galaxy properties. Specifically, it has the largest ${\rm EW}_{\rm Ly\alpha}$ ($\sim43\,$\AA\ as compared to $\sim13\,$\AA\ in non LyC-emitting regions) and $f_{\rm esc}^{\rm Ly\alpha}$ ($\sim30\%$, as compared to $13\%$ in non LyC-emitting regions) in the galaxy \citep{Kim2023}. 

For other galaxies with resolved LyC and \lya\ measurements, the picture is not as simple. In nearby galaxy Haro 11, the star-forming region with highest $f_{\rm esc}^{\rm Ly\alpha}$ shows higher $f_{\rm esc}^{\rm LyC}$, however, the LyC luminosity is almost double in the knot with the lowest $f_{\rm esc}^{\rm Ly\alpha}$ \citep{Komarova2024}. Additionally, \lya\ and LyC observations of the LyC-emitter with highest $f_{\rm esc}^{\rm LyC}$ ($\sim90$\%) currently known at $z\sim3.6$ show weak \lya\ emission in the LyC-emitting region \citep{Marques-Chaves2024}. This is expected in a case where nearly all of the LyC escape, indicating a very low \hi\ content, and thus, little material for \lya\ production, on the line-of-sight to the starburst. With only a few LyC-emitting galaxies with sub-kpc \lya\ measurements, all occupying different redshift ranges, the link between resolved \lya\ and LyC emission is currently not well established.

The resolution reached by LaCOS (with a PSF FWHM corresponding to $\sim$400 pc) enables studies of the impact of resolved galaxy properties on the LyC escape fraction in a larger galaxy sample than possible before. However, resolved LyC measurements cannot currently be obtained with available instrumentation. Thus, in the following, we  measure the impact of resolved properties onto the line-of-sight $f_{\rm esc}^{\rm LyC}$ integrated over the COS aperture. Specifically, we investigate possible correlations between the LyC escape fraction and resolved \lya\ and UV continuum observables. We also note that without \ha\ photometry currently available, the sub-kpc \lya\ escape fraction cannot be derived.
\begin{figure*}[t]
    \centering
    \includegraphics[width=0.7\textwidth]{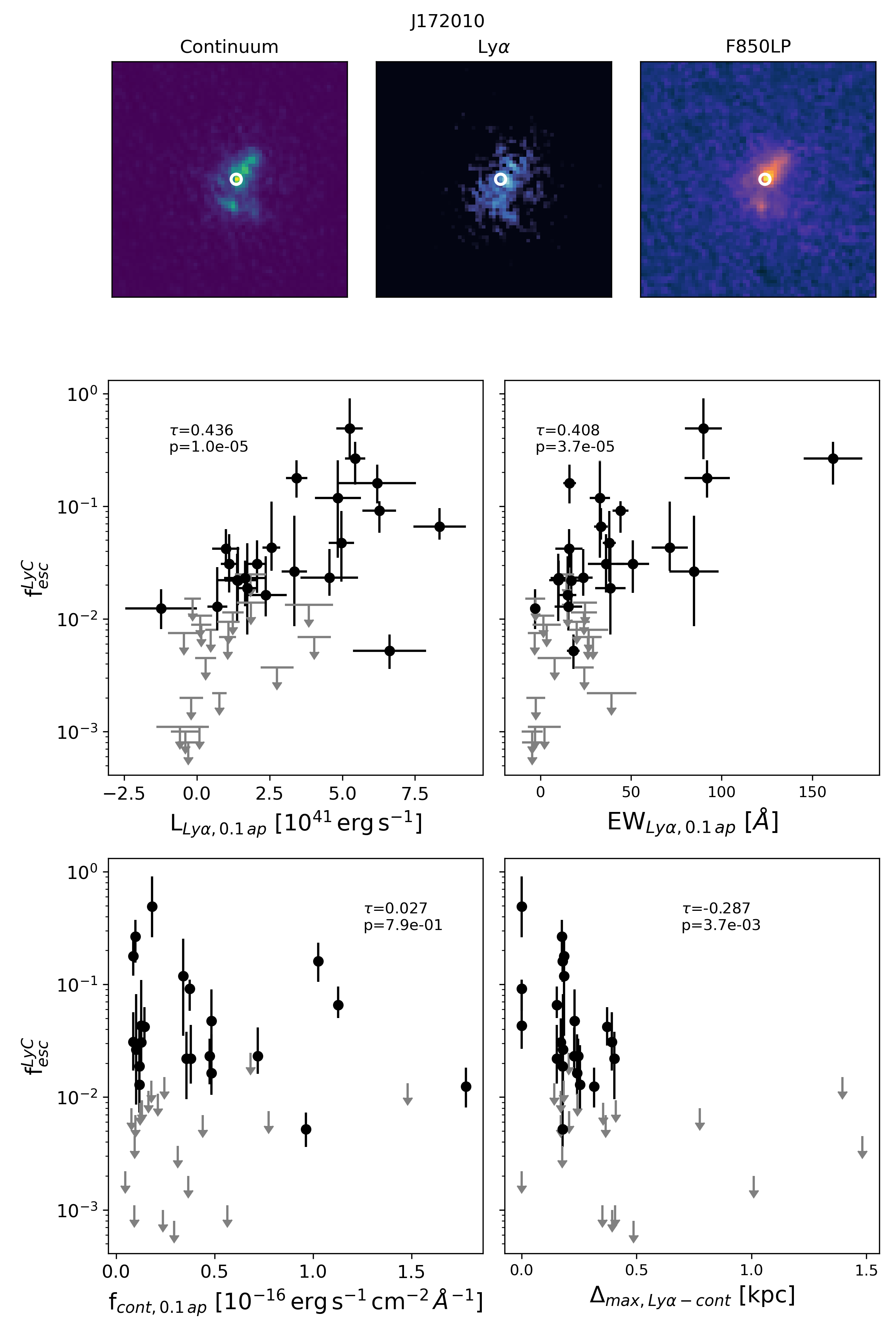}
    \caption{Line-of-sight LyC escape fraction measured in the COS aperture as a function of observables extracted in a 0.12\arcsec\ aperture centered around the brightest UV continuum pixel in each galaxy. The first top row of panels shows the position of the aperture on the UV continuum (left), \lya\ (middle) and F850LP (right) images of galaxy J172010 as an example. The panels below show on their $x$-axis, from top to bottom and left to right, the \lya\ luminosity, EW, the UV continuum flux density and the spatial offset between the brightest UV continuum pixel and the brightest \lya\ pixel as a function of $f_{\rm esc}^{\rm LyC}$. We show the Kendall $\tau$ and associated $p$-value assessing the degree of correlation between variables on the top corners of the plots. In general, there is a high degree of correlation between the \lya\ observables within the aperture and the LyC escape fraction.} 
    \label{fig:lya_res_maxpixuv}
\end{figure*}

To investigate the impact of sub-kpc scale \lya\ properties on $f_{\rm esc}^{\rm LyC}$, we specifically derive \lya\ and UV continuum observables within a small aperture (3 pixels in diameter, corresponding to 0.12\arcsec, roughly the PSF FWHM) centered on the brightest UV continuum sources in each galaxy. Figure \ref{fig:lya_res_maxpixuv} shows scatter plots presenting $f_{\rm esc}^{\rm LyC}$ as a function of L$_{\rm Ly\alpha}$, EW$_{\rm Ly\alpha}$, $f_{\rm \lambda,UV}$ obtained in this aperture as well as $\Delta_{\rm max,\, Ly\alpha-cont}$, the spatial offset between the brightest UV continuum and \lya\ pixel. The Figure also shows UV continuum, \lya\ and F850LP images of J172010 with the aperture overlaid  as an example.

Inspecting the scatter plots in Figure \ref{fig:lya_res_maxpixuv}, We do not find any indication of correlation between the UV continuum flux in the brightest UV-continuum region and $f_{\rm esc}^{\rm LyC}$ ($\tau=0.02,\ p=0.8$). However, we find excellent correlations between both the $L_{\rm Ly\alpha}$ ($\tau=0.44,\ p=1e-5$) and EW$_{\rm Ly\alpha}$ ($\tau=0.41,\ p=3.7e-5$), and $f_{\rm esc}^{\rm LyC}$, as well as a tentative anti-correlation between $f_{\rm esc}^{\rm LyC}$ and $\Delta_{\rm max,\, Ly\alpha-cont}$ ($\tau=-0.29,\ p=3.7e-3$). The correlation between \lya\ observables and $f_{\rm esc}^{\rm LyC}$ is stronger when evaluated in small apertures, rather than when measured in the COS aperture or via global photometry. These results indicate that the brightest LyC-emitting regions in a galaxy may contribute significantly to the escaping LyC budget on the line-of-sight. Specifically, the presence of neutral gas in front of these sources would result in lower, or negative $L_{\rm Ly\alpha}$ and ${\rm EW}_{\rm Ly\alpha}$, and larger offsets between the peak \lya\ and UV continuum pixels. Thus, we interpret the trends observed in Figure \ref{fig:lya_res_maxpixuv} as indication that a few compact, bright and unobscured UV-continuum sources might drive the observed LyC escape in LaCOS galaxies. 

\subsection{LyC escape and compactness}
Small galaxy sizes and high star formation rate surface density $\Sigma_{\rm{\rm SFR}}$, have been linked to increased LyC escape fractions \citep{Heckman2001,Marchi2018,Naidu2020,Cen2020,Yeh2023,Marques-Chaves2024}. The star formation rate surface density is a function of the half-light radius $r_{50}$, defined as:
\begin{equation}
 \Sigma_{\rm{\rm SFR}}=  \frac{\rm SFR}{2\,\pi\,r_{50}^2} \, .
 \label{eq:SSFR}
\end{equation}
It has been suggested as a probe of star formation feedback, a mechanism thought to be responsible for clearing the neutral ISM around ionizing sources and facilitating LyC escape \citep[e.g.][]{Jaskot2017,Gazagnes2020,Komarova2021,Amorin2024,Carr2024,Flury2024}. Both decreased $r_{50}$ and increased $\Sigma_{\rm{\rm SFR}}$ have been found to correlate with $f_{\rm esc}^{\rm LyC}$ in the LzLCS sample \citep{Flury2022b}. However, previous measurements of the UV half light radius for LzLCS galaxies were obtained from relatively shallow and vignetted COS acquisition images. This could potentially bias measurements and affect correlations between $r_{50}$, $\Sigma_{\rm{\rm SFR}}$ and the escape fraction, especially in the more extended galaxies. 

\begin{figure}[h]
    \centering
    \includegraphics[width=0.99\linewidth]{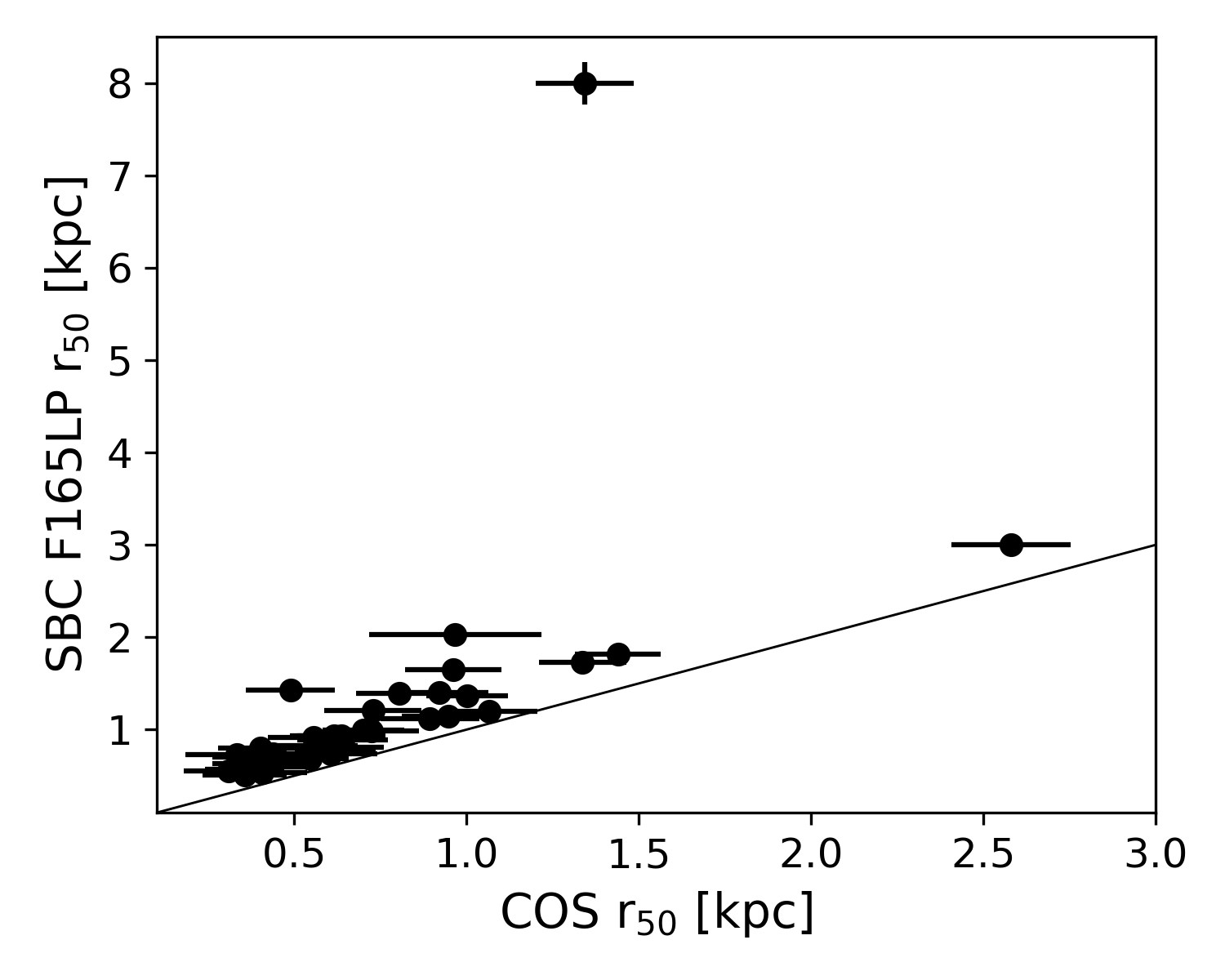}
    \caption{Comparison between UV half-light radius calculated using COS acquisition images and SBC photometry. A solid black line indicates the 1:1 relation. The $r_{50}$ values obtained with SBC are slightly larger, but in agreement with COS measurements, except for galaxy J095700.}
    \label{fig:r50_comp}
\end{figure}

Here, we re-derive the half light radius $r_{50}$ previously measured from COS acquisition images, using instead the deeper SBC F165LP photometry. To do so, we integrate the F165LP light profiles in increasingly larger circular apertures centered on the brightest pixel in the image. We use a 0.5 pixel increment, and compute the flux as a function of radius. We determine the flux density by integrating within a circular aperture at which the flux reaches a plateau (less than $10^{-3}$ fractional flux difference). Then, $r_{50}$ is determined from the same profile as the radius at which the galaxy reaches half of its flux density in the F165LP image. The comparison between COS acquisition images and SBC frames $r_{50}$ is presented in Figure \ref{fig:r50_comp}. The SBC $r_{50}$ values are systematically larger than the COS ones. In most cases, the difference is relatively small, with the SBC $r_{50}$ being larger by an average of $0.5\,\rm{kpc}$, or $0.1\,$\arcsec. One galaxy is a significant outlier, with a SBC $r_{50}$ that is 6 times that of the COS value. This is J095700, the largest object in the sample (see Appendix Figure \ref{fig:lacos-large_gals}), that appears to be an actively merging system. This galaxy is large both in the rest-frame optical and UV, but the bulk of the UV emission is concentrated in a clump in the center. The fainter, extended UV emission was likely missed in the COS acquisition image since it is shallow and vignetted. However, this galaxy constitutes the main exception to the relatively small offset between $r_{50}$ derived from COS and SBC. This indicates that, as expected, the deeper and non-vignetted SBC photometry does recover light from extended regions, resulting in the galaxies being slightly larger than found in previous LzLCS studies. Nevertheless, with a characteristic median $r_{50}=0.8\,$ kpc, the galaxies are still very compact.

\begin{figure}
    \centering
    \includegraphics[width=0.99\linewidth]{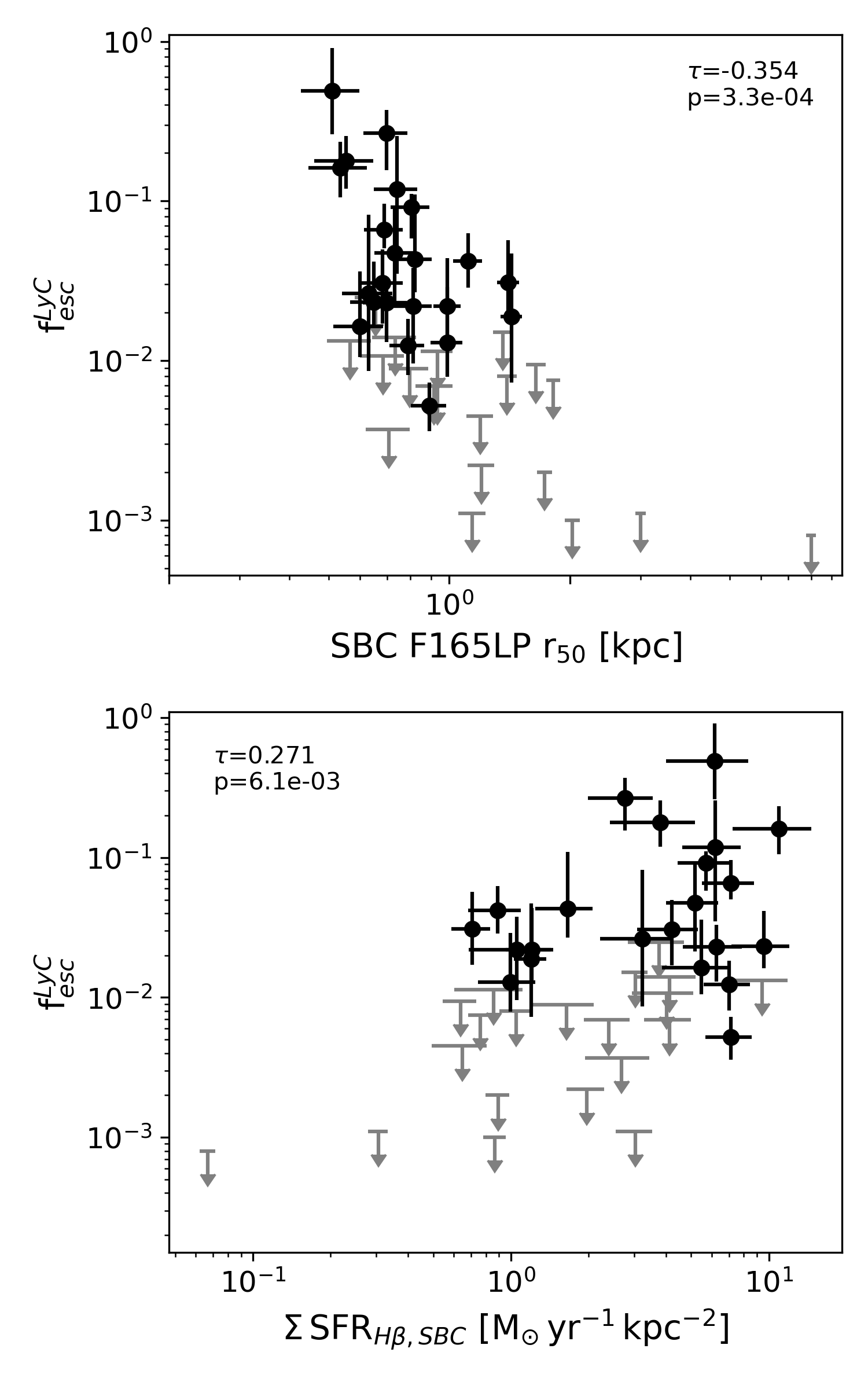}
    \caption{Top panel: LyC escape fraction as a function of $r_{50}$ obtained from F165LP SBC images. Bottom panel: LyC escape fraction as a function of $\Sigma_{\rm{\rm SFR}}$. The trends with $f_{\rm esc}^{\rm LyC}$ are the same as those measured with spectroscopy, but the correlations are weaker.}
    \label{fig:r50_fesc}
\end{figure}

We next look at the link between the newly derived half-light radius $r_{50}$ and the LyC escape fraction. We additionally re-derive $\Sigma_{\rm{\rm SFR}}$ using the new $r_{50}$ values and the H$\beta$ SFR following Eq.~(\ref{eq:SSFR}).
Plots showing the escape fraction as a function $r_{50}$ and $\Sigma_{\rm{\rm SFR}}$ are shown  on the top and bottom panels of Figure \ref{fig:r50_fesc}, respectively. 
Generally, we find similar trends with the newly derived photometric values and the spectroscopic measurement. The newly obtained $\tau$ values are within the typical uncertainties on $\tau\sim0.05$ in spectroscopy \citep{Flury2022b} (see Table \ref{tab:correlations_COSvSBC}). Thus, the improved precision on $r_{50}$ still indicates that concentrated star formation is key to LyC escape. The escaping LyC emission is likely to originate from small, compact and bright star-forming regions that create the conditions for increased stellar feedback, rather than diffuse ones. This echoes the results on the strong correlation between sub-kpc \lya\ observables and $f_{\rm esc}^{\rm LyC}$.
\begin{table}[h]
    \footnotesize
    \centering
    \caption{Strength of the correlations between the LyC escape fraction and parameters measured using either COS or SBC data.}
    \begin{tabular}{ccccccc}
    \hline
 par vs $f_{\rm esc}$ & \multicolumn{2}{c}{SBC$_{\rm tot}$\tablenotemark{a}}& \multicolumn{2}{c}{SBC$_{\rm 0.1\arcsec\,ap}$\tablenotemark{b}} & \multicolumn{2}{c}{COS\tablenotemark{c}} \\ 
  &$\tau$&$p$ &$\tau$&$p$ &$\tau$&$p$ \\
\hline
 $L_{\rm Ly\alpha}$ & 0.25 & 1.2e-2 & \textbf{0.44} & \textbf{1e-5} & \textbf{0.37} & \textbf{1.8e-4} \\ 
 EW$_{\rm Ly\alpha}$  & \textbf{0.36} & \textbf{2.8e-4} & \textbf{0.41} & \textbf{3.7e-5} & \textbf{0.38} & \textbf{1.1e-4} \\ 
 $f_{\rm esc}^{\rm Ly\alpha}$  & 0.29 & 3.7e-3 & & & \textbf{0.43} & \textbf{1.6e-5}\\ 
 r$_{50}$ & $\mathbf{-0.35}$ & \textbf{3.3e-04} & & & $\mathbf{-0.38}$ &\textbf{9.9e-5}  \\ 
 $\Sigma_{\rm{\rm SFR}}$  & 0.27 & 6.1e-3 & & & 0.24& 1.5e-2 \\ 
 \hline
    \end{tabular}
    \tablecomments{The Kendall $\tau$ and $p$-value are calculated on matched samples for the different measurements. Consistent with LzLCS studies, a correlation is deemed significant if $p<0.00135$, consistent with previous LzLCS studies, in which case the values are shown in bold.}\tablenotetext{a}{Values measured in a large aperture in SBC photometry.}\tablenotetext{b}{Values measured in SBC photmetry in a $0.1\,$\arcsec\ aperture centered on the brightest UV continuum pixel.}\tablenotetext{c}{Values measured in the COS aperture.}
    \label{tab:correlations_COSvSBC}
\end{table}

\section{Summary}
\label{sec:conclusion}
We have presented HST photometric data and data products for the LaCOS sample, a subset of 42 low-$z$ LyC-emitting and non-emitting galaxies representative of galaxies in the LzLCS. LaCOS observations include data obtained in three rest-frame optical bands and in two  rest-frame UV bands, sampling the stellar continuum and \lya\ line. We present optical RBG composite, \lya\ flux and EW maps for all galaxies in the sample. The LaCOS objects show a wide range of morphologies, but are mostly irregular galaxies, both for the LyC leakers and non-leakers. Additionally, they have a variety of \lya\ flux and EW morphologies and distributions. Forthcoming manuscripts will specifically investigate the impact of galaxy morphology, merger interaction, and \lya\ halo extent on LyC production and escape.

With the newly obtained HST photometry, we re-investigate correlations previously examined with spectroscopic data between LyC escape fraction and global galaxy properties. Specifically, we re-derive the \lya\ luminosity, equivalent width, the \lya\ escape fraction, the half-light radius and the star formation rate surface density.
We find similar trends with $f_{\rm esc}^{\rm LyC}$ for measurements derived through photometry as those previously found with spectroscopic measurements.  Generally, the trends with $f_{\rm esc}^{\rm LyC}$ show a weaker degree of correlation when obtained with global photometry rather than spectroscopy, which could be due to the uncertainties on photometric values, and to photometry recovering light from extended regions that do not contribute significantly to the escaping LyC photon output on the line of sight. With global photometry, we confirm a robust correlation ($p<0.00135$) between $f_{\rm esc}^{\rm LyC}$ and EW$_{\rm Ly\alpha}$, a robust anti-correlation between $f_{\rm esc}^{\rm LyC}$ and $r_{50}$, and find tentative correlations ($p<0.05$) between $f_{\rm esc}^{\rm LyC}$ and $L_{\rm Ly\alpha}$, $f_{\rm esc}^{\rm Ly\alpha}$, and $\Sigma_{\rm{\rm SFR}}$. 

Finally, we investigate possible correlations between sub-kpc galaxy properties and line-of-sight $f_{\rm esc}^{\rm LyC}$ derived from the COS spectra. Specifically, we measure \lya\ and UV continuum in small apertures around the brightest UV continuum source in each galaxy. While we cannot currently investigate correlations with $f_{\rm esc}^{\rm Ly\alpha}$, we find strong correlations between the line-of-sight $f_{\rm esc}^{\rm LyC}$ and the \lya\ luminosity and EW. The correlations with $L_{\rm Ly\alpha}$ and ${\rm EW}_{\rm Ly\alpha}$ are much stronger when investigated on sub-kpc scales ($\sim400\,$pc), than on larger scales with photometry or spectroscopy. Additionally, we do not find any correlation between the continuum flux density around the brightest continuum source and $f_{\rm esc}^{\rm LyC}$, and find a tentative correlation ($p=0.0037$) between the spatial offset between peak UV continuum and peak \lya\ emitting  pixels. We interpret these result as evidence that the brightest UV continuum sources in LaCOS galaxies likely contribute a large fraction of the escaping LyC radiation on the line of sight, and that the lack of neutral gas in front of these sources (as traced by smaller \lya-to-UV offsets and higher $L_{\rm Ly\alpha}$ and EW$_{\rm Ly\alpha}$) likely plays a determining role on the line-of-sight $f_{\rm esc}^{\rm LyC}$.

\begin{acknowledgments}
This research is based on observations made with the NASA/ESA Hubble Space Telescope obtained from the Space Telescope Science Institute, which is operated by the Association of Universities for Research in Astronomy, Inc., under NASA contract NAS 5–26555. These observations are from HST GO programs 17069, 14131 and 11107. ALR acknowledges support from  HST GO-17069. MJH is supported by the Swedish Research Council (Vetenskapsr{\aa}det) and is fellow of the Knut and Alice Wallenberg Foundation. ASL acknowledges support from the Knut and Alice Wallenberg Foundation. SB acknowledges support from NSF grant 2408050.
FL acknowledges funding from the European Union's Horizon 2020 research and innovation program under the Marie Skodowska-Curie grant agreement No. C3UBES-101107619. 
\end{acknowledgments}

\software{numpy \citep{numpy}, astropy \citep{astropy1,astropy2,astropy3}, matplotlib \citep{matplotlib}, spectres \citep{spectres}, photutils \citep{photutils}, dust\_extinction \citep{dustext}, astroscrappy \citep{vanDokkum2001,astroscrappy}, CROCOA \citep{crocoa}, scipy \citep{scipy}, cmocean \citep{cmocean}, cmastro.}

\bibliography{bibliography}{}
\bibliographystyle{aasjournal}

\appendix

\section{Observation parameters}
In Table \ref{tab:hst_obs} we present the parameters for the HST  observations of LaCOS galaxies. In particular, we specify the exposure time per filter for each galaxy, and the temperature range of the detectors for SBC filters.
\begin{table*}[t]
    \centering
    \caption{List of exposure times and detector temperatures. The exposure times are given for the different HST filters used to observe the LaCOS galaxies. The detector temperature ranges are shown for the two ACS/SBC filters, F150LP and F165LP.}
    \begin{tabular}{cccccccc}
\hline
ID & $t_{\rm F150LP}$[s] & $t_{\rm F165LP}$[s] & $t_{\rm F438W}$[s] & $t_{\rm F547M}$[s] & $t_{\rm F850LP}$[s] & $T_{\rm F150LP}$[$^{\circ}$C] & $T_{\rm F165LP}$[$^{\circ}$C] \\ 
\hline
J011309  &  2000  &  2568  &  508  &  613  &  696  &  17.9 - 25.1  &  18.9 - 24.8  \\ 
J012910  &  2000  &  2586  &  508  &  622  &  696  &  18.7 - 25.6  &  19.6 - 25.3  \\ 
J072326  &  2000  &  2590  &  508  &  624  &  696  &  18.7 - 25.5  &  19.6 - 25.1  \\ 
J081409  &  2000  &  2590  &  508  &  624  &  696  &  18.9 - 26.1  &  20.1 - 25.8  \\ 
J082652  &  1984  &  2590  &  508  &  624  &  696  &  17.7 - 25.6  &  18.8 - 25.3  \\ 
J090918  &  1980  &  1947  &  508  &  622  &  696  &  18.4 - 25.9  &  19.4 - 25.5  \\ 
J091113  &  1000  &  1311  &  508  &  624  &  696  &  24.8 - 26.0  &  25.3 - 25.6  \\ 
J091207  &  2000  &  2616  &  508  &  637  &  696  &  18.2 - 26.6  &  19.2 - 26.2  \\ 
J091703  &  2000  &  2586  &  508  &  622  &  696  &  17.7 - 24.7  &  18.7 - 24.4  \\ 
J092532  &  2244  &  6133  &  900  &  1499  &  776  &  17.8 - 25.5  &  19.9 - 24.7  \\ 
J092552  &  2000  &  2586  &  508  &  622  &  696  &  18.2 - 25.4  &  19.4 - 24.9  \\ 
J093355  &  1962  &  2610  &  508  &  637  &  696  &  18.2 - 25.1  &  19.2 - 24.7  \\ 
J095236  &  2000  &  2590  &  508  &  624  &  696  &  19.0 - 26.2  &  20.1 - 25.9  \\ 
J095700  &  2000  &  2590  &  508  &  624  &  696  &  17.4 - 24.7  &  18.4 - 24.3  \\ 
J095838  &  2000  &  2590  &  508  &  624  &  696  &  18.2 - 25.4  &  19.2 - 25.0  \\ 
J105331  &  2000  &  2616  &  508  &  637  &  696  &  18.9 - 26.1  &  19.9 - 25.8  \\ 
J110452  &  2000  &  2598  &  508  &  628  &  696  &  18.2 - 25.4  &  19.4 - 25.0  \\ 
J112224  &  1485  &  1919  &  508  &  624  &  696  &  18.2 - 23.8  &  19.2 - 24.3  \\ 
J113304  &  2715  &  2323  &  508  &  665  &  696  &  13.8 - 16.3  &  18.7 - 21.5  \\ 
J115855  &  2000  &  2586  &  508  &  622  &  696  &  20.6 - 27.6  &  21.6 - 27.3  \\ 
J115959  &  2000  &  2586  &  508  &  622  &  696  &  18.4 - 25.6  &  19.5 - 25.3  \\ 
J120934  &  2000  &  2586  &  508  &  622  &  696  &  18.9 - 25.9  &  19.9 - 25.6  \\ 
J121915  &  2000  &  2598  &  508  &  628  &  696  &  27.1 - 29.3  &  27.3 - 29.2  \\ 
J124033  &  2000  &  2590  &  508  &  624  &  696  &  17.9 - 24.9  &  19.2 - 24.5  \\ 
J124423  &  2000  &  2568  &  508  &  613  &  696  &  18.4 - 25.6  &  19.4 - 25.3  \\ 
J124835  &  3080  &  8160  &  900  &  1499  &  764  &  17.3 - 24.5  &  19.4 - 24.7  \\ 
J125503  &  2000  &  2588  &  508  &  623  &  696  &  18.5 - 25.6  &  19.6 - 25.3  \\ 
J125718  &  2000  &  2590  &  508  &  624  &  696  &  18.9 - 25.9  &  19.9 - 25.6  \\ 
J130559  &  1978  &  2568  &  508  &  624  &  696  &  18.1 - 24.9  &  19.2 - 24.5  \\ 
J131037  &  2000  &  2590  &  508  &  624  &  696  &  18.7 - 25.6  &  19.6 - 25.3  \\ 
J131419  &  2000  &  2586  &  508  &  622  &  696  &  18.4 - 25.6  &  19.6 - 25.3  \\ 
J131904  &  1984  &  2584  &  508  &  637  &  696  &  18.9 - 21.8  &  19.9 - 21.1  \\ 
J132633  &  1984  &  2558  &  508  &  624  &  696  &  18.2 - 21.1  &  19.4 - 20.5  \\ 
J132937  &  2000  &  2632  &  508  &  645  &  696  &  17.9 - 25.0  &  19.2 - 24.7  \\ 
J134559  &  2000  &  2586  &  508  &  622  &  696  &  18.7 - 25.9  &  19.9 - 25.6  \\ 
J140333  &  2000  &  2648  &  508  &  653  &  696  &  18.4 - 25.5  &  19.4 - 25.1  \\ 
J144010  &  2000  &  2598  &  508  &  628  &  696  &  18.4 - 25.5  &  19.4 - 25.1  \\ 
J154050  &  2000  &  2632  &  508  &  645  &  696  &  17.8 - 24.2  &  18.9 - 23.8  \\ 
J155945  &  2000  &  2590  &  508  &  624  &  696  &  18.8 - 25.6  &  19.9 - 25.4  \\ 
J160437  &  2000  &  2576  &  508  &  617  &  696  &  18.4 - 25.6  &  19.6 - 25.3  \\ 
J164607  &  2000  &  2586  &  508  &  622  &  696  &  18.4 - 27.0  &  19.6 - 26.8  \\ 
J172010  &  2000  &  2616  &  508  &  637  &  696  &  19.2 - 26.1  &  20.1 - 25.9  \\ 

\hline
    \end{tabular}
    \label{tab:hst_obs}
\end{table*}

\section{COS aperture}

Obtaining accurate \lya\ maps with the method presented in this manuscript requires precisely matching the photometric aperture used to calculate the bandpass normalization factor $\alpha$ to the actual COS aperture. We initially consider an aperture taking into account the vignetting of the COS aperture and the mask used to extract high S/N \lya\ spectra in \citet{Flury2022a} (hereafter, aperture 1). This extraction mask corresponds to a slit of 0.637\arcsec\ width oriented along the COS cross-dispersion axis. However, using aperture 1 results in significant offset \lya\ fluxes between HST photometry and COS spectroscopy, with spectroscopic fluxes being systematically higher on the order of 75\%. We also compare the photometric flux extracted in the COS aperture for the individual UV filters, F150LP and F165LP, to synthetic photometry obtained with the COS spectra. Depending on the filter considered, we find that fluxes obtained from the COS spectra are systematically larger by 40 to 45\%, which cannot be accounted for by the flux uncertainty on COS or SBC ($<5$\%). 

\begin{figure}
    \centering
    \includegraphics[width=\textwidth]{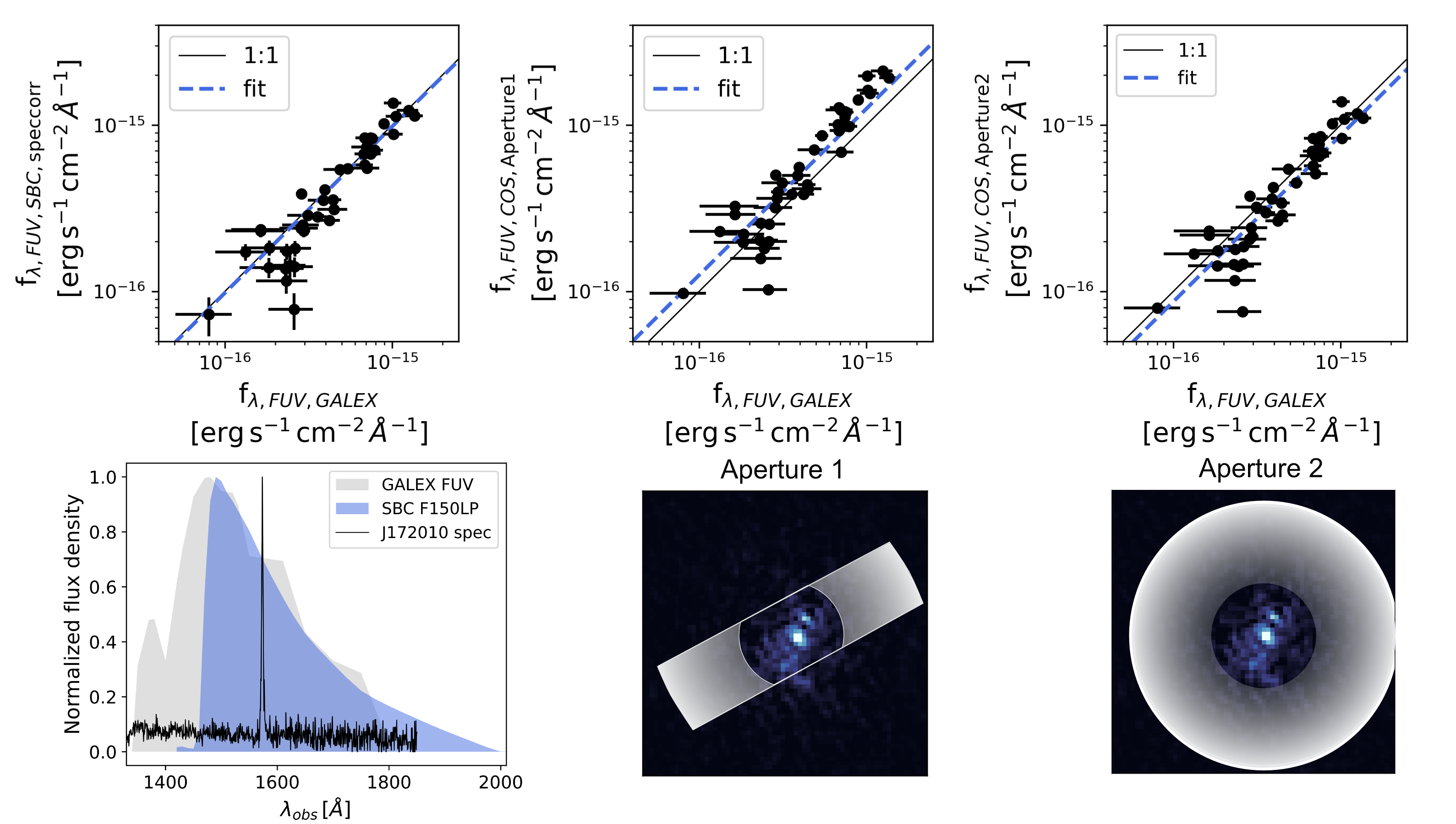}
    \caption{Comparison of GALEX, SBC and COS fluxes. The top left panel shows the bandpass-corrected SBC F150LP fluxes extracted in a 6\arcsec\ radius aperture, similar to the GALEX one. The two other top panels show the fluxes from COS synthetic photometry, corrected using F150LP photometry (due to the small sizes of the apertures) to match the value that would be measured in the GALEX aperture. The middle top panel shows what would be measured with an  aperture combining the COS vignetting profile and LzLCS spectral extraction window (aperture 1), and the right top panel, the circular 2.5\arcsec\ diameter aperture value (aperture 2). Blue lines show linear fits to the data.
    The bottom panels show, from left to right, the normalized bandpass response of GALEX FUV and SBC F150LP overlaid the spectrum for J172010 spectra, and the two apertures used to correct the COS synthetic FUV photometry, overlaid on the \lya\ image of J172010.}
    \label{fig:GALEXvCOSvSBC}
\end{figure}

To assess whether the offset could be due to flux calibration issues with SBC or COS data, we compare these measurements to independent estimates of the FUV flux obtained from GALEX (shown in Figure \ref{fig:GALEXvCOSvSBC}), since the SBC F150LP and GALEX FUV filter overlap considerably (see bottom left panel of the Figure). We compare GALEX FUV fluxes with the SBC F150LP fluxes obtained within a 6\arcsec\ aperture in radius. While the FUV and F150LP filter coverage are very similar, their bandpasses differ slightly. Thus, to enable comparison, we correct the SBC fluxes for the difference in bandpass coverage using the available COS spectra. Specifically, we perform synthetic photometry on the COS spectra to calculate the flux that would be measured respectively with the SBC F150LP and GALEX FUV filters for a given galaxy. The correction factor applied to SBC fluxes is the ratio of the synthetic photometry fluxes, ranging from 0.89 to 1.06. We find a relatively good agreement between the GALEX and bandpass-corrected SBC fluxes, with most data points hovering around the 1:1 line, as can be seen on the top left panel of Figure \ref{fig:GALEXvCOSvSBC}. Therefore, we conclude that the offset between COS and SBC fluxes is not due to SBC flux calibration issues.

Next, we compare the COS synthetic photometry fluxes to those obtained with GALEX. Since the COS aperture is significantly smaller than that of GALEX (at most, a circular aperture with 1.25\arcsec\ radius, as opposed to a 6\arcsec\ radius aperture) and to make a comparison possible, we correct COS fluxes using the SBC images. To obtain correction factors, we calculate the ratio of fluxes computed within the COS and GALEX apertures in the SBC F150LP images. The correction factors range from 1.62 to 5.10. We find a strong disagreement between COS and GALEX fluxes when using aperture 1, with COS fluxes being higher than GALEX values by $\sim$60\% (see middle panel of Figure \ref{fig:GALEXvCOSvSBC}). 

 To further investigate whether the discrepancy is caused by the assumption on shape aperture, we test the effect of using the simpler vignetted circular aperture that is 2.5\arcsec\ in diameter instead (Hereafter, aperture 2, shown on the bottom right panel of Figure \ref{fig:GALEXvCOSvSBC}). When repeating the analysis outlined above with this aperture, we find a better agreement between GALEX and COS fluxes, with COS fluxes being 10\% lower than GALEX on average, as demonstrated in the top right panel of Figure \ref{fig:GALEXvCOSvSBC}. Additionally, this aperture solves the disagreement between COS and SBC F150LP and F165LP fluxes within the aperture. 
 The fact that this aperture leads to better agreement with photometry as opposed to aperture 1 could be caused by optics smearing the target flux in the cross-dispersion direction (private communication with STScI). Thus, we choose to conduct subsequent analysis and to derive the \lya\ maps using aperture 2. 
\label{app:COSap}

\section{Comparison to LaXs}

\label{app:LaXs}
In the low-z Universe ($z\sim0.03$), the pixel-wise SED fitting software LaXs \citep{Hayes2009} has been commonly used to obtain \lya\ maps of star-forming galaxies \citep{Hayes2014,Melinder2023}. Most galaxies in LaCOS do not have the number of filters required for robust pixel-by-pixel fitting of the data using LaXs. However, two galaxies in the sample were observed as part of archival HST programs containing additional filters. In particular, J12483 was observed as part of HST program 14131, and J092532 as part of program 14466, both with additional ACS/WFC observations in filter FR853N covering the \ha\ line. This photometric coverage allows for robust pixel-by-pixel fitting of the data, allowing for \lya\ maps to be produced with LaXs. In this section, we compare the \lya\ and UV continuum maps obtained for J124835 and J092532 using the method devised for the LaCOS sample with those obtained from photometric fitting using LaXs. To rule out any possible differences coming from different data reduction methods, we apply the methods used for the LaCOS sample to the science frames reduced by collaborators for J124835 and J092532 and used in the LaXs run (described in detail in Melinder et al., in prep.).

\begin{figure}
    \centering
    \includegraphics[width=\textwidth]{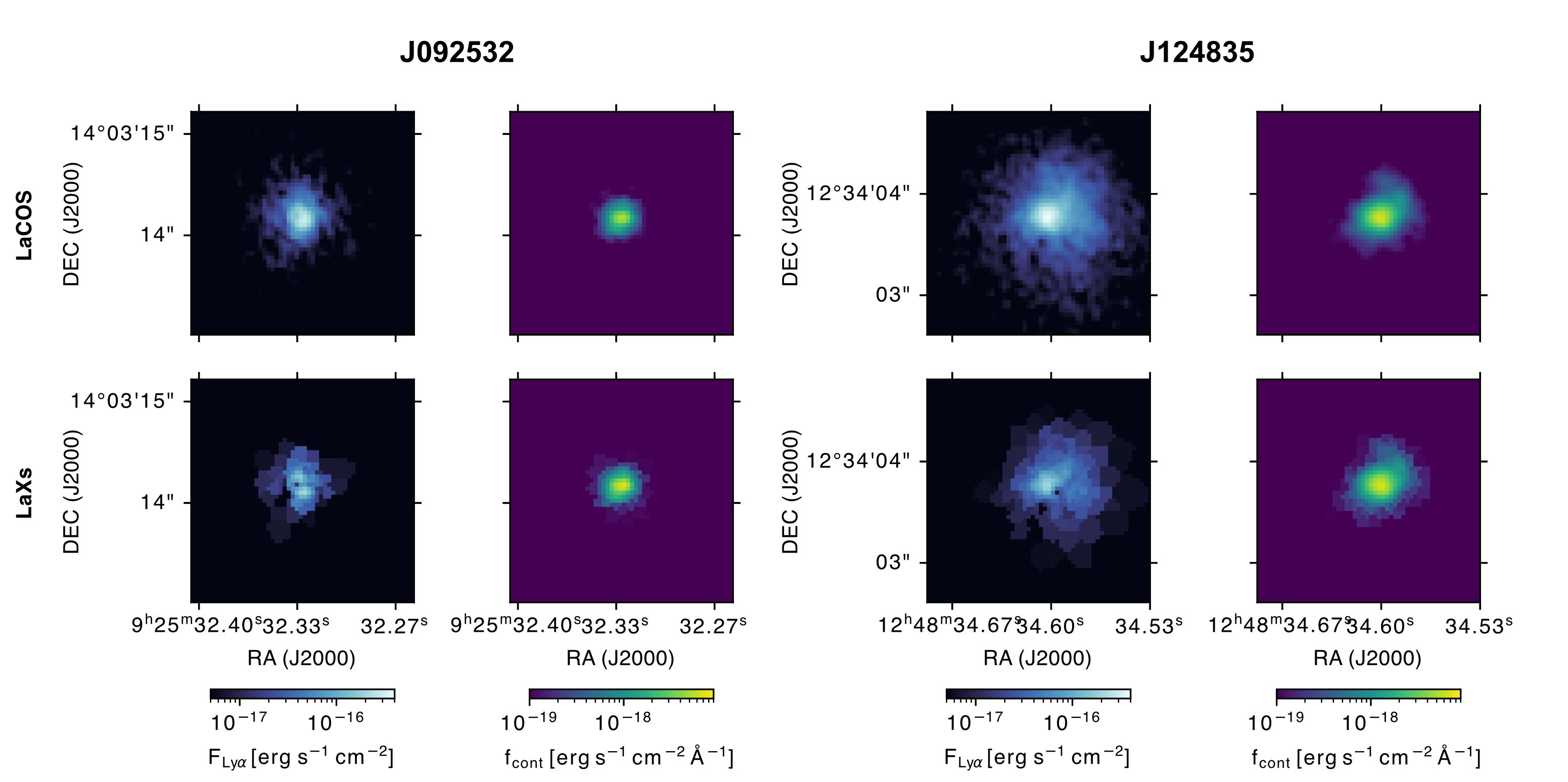}
    \caption{Comparison between LaCOS (top panels) and LaXs (bottom pannels) \lya\ and UV continuum maps for J092532 (left set of panels) and J124835 (right set of panels).}
    \label{fig:test_LaXs}
\end{figure}

The maps showing LaCOS and LaXs \lya\ and UV continuum maps for J092532 and J124835 are shown in Figure \ref{fig:test_LaXs}. The UV continuum maps are extremely similar for both derivations, showing the same morphology but with slightly brighter emission in LaXs maps. However, the LaXs \lya\ maps are much fainter than the LaCOS ones, with a similar global morphology, but additional \lya\ absorption regions in LaXs. Table \ref{tab:test_LaXs} presents the \lya\ flux, continuum flux density and \lya\ EW values integrated within the COS aperture for the two galaxies using maps produced. By construct, the equivalent width within the COS aperture is the same for the LaCOS map and the COS spectrum. However, the LaXs-derived EW value in the COS aperture is smaller by a significant amount. The UV continuum flux density values are smaller (with average 17\% relative difference) for LaCOS as compared to LaXs map. The \lya\ flux values in the photometric maps are smaller than those derived from COS spectroscopy, however the LaXs value is smaller by a larger amount than the LaCOS maps (with average 34\% relative difference between LaCOS and LaXs). This indicates that, at least within the COS extraction aperture, the LaCOS maps are in better agreement with the values from spectroscopy, which is expected given the method employed to derive the maps.

The \lya\ measurements from the maps derived using LaXs are significantly different from the spectroscopic values. This behavior has previously been noted with LaXs, and may be due to continuum emission being over-estimated by the software in regions with low signal-to-noise. It is beyond the scope of this study to investigate and correct for this offset, but possible causes could be uncertainties in the absolute calibration of HST instruments, or assumptions underpinning calculations in LaXs. However, we note this could impact comparison of \lya\ observables in high-z studies and those in the very nearby Universe, where LaXs has been employed \citep[e.g.][]{Melinder2023}.

\begin{table}[h]
    \centering
    \caption{Comparison between \lya\ and UV continuum measurements obtained in the COS aperture for J092532 and J124835, using the maps derived using the LaCOS method and LaXs, and the values calculated from the COS spectrum.}
    \begin{tabular}{c|ccc|ccc}
    \hline
        & \multicolumn{3}{c|}{J092538} & \multicolumn{3}{c}{J124835}  \\
    \hline
        & $F_{\rm Ly\alpha}$ & $f_{\rm cont}$& EW &
        $F_{\rm Ly\alpha}$ & $f_{\rm cont}$& EW \\
        & $10^{-14}$ erg$\,$s$^{-1}\,$cm$^{-2}$ & $10^{-16}$ erg$\,$s$^{-1}\,$cm$^{-2}\,$\AA$^{-1}$ & \AA &
        $10^{-14}$ erg$\,$s$^{-1}\,$cm$^{-2}$ & $10^{-16}$ erg$\,$s$^{-1}\,$cm$^{-2}\,$\AA$^{-1}$ & \AA \\
    \hline
    LaCOS & $1.48 \pm 0.04$ & $1.53 \pm 0.16$ & $75 \pm 8$ 
 & $3.89\pm0.05$ & $3.16\pm0.26$ & $97\pm8$\\
    LaXs & $1.13\pm0.04$ & $1.93\pm0.02$ &  $45\pm2$ & $2.17\pm0.03$ & $3.42\pm0.02$ & $50 \pm 1$  \\
    COS & $2.20\pm0.09$ & - & $75\pm6$ & $5.78\pm0.14$ & - & $97\pm8$
\\ 
    \hline
    \end{tabular}
    \label{tab:test_LaXs}
\end{table}

\section{Figures for large galaxies}
\begin{figure*}[h]
    \centering
    \includegraphics[width=0.80\textwidth]{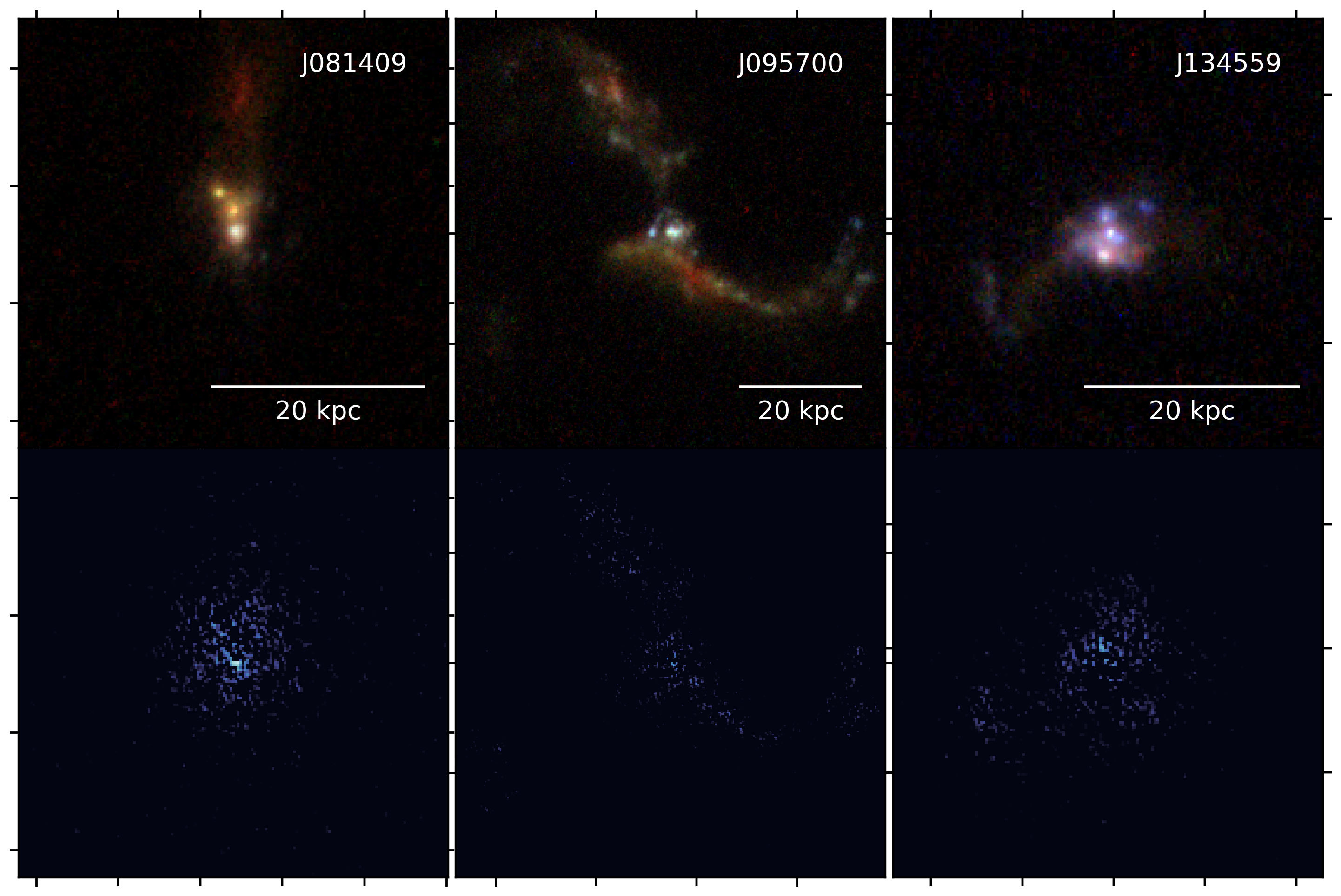}
    \caption{Optical RGB composite (top panels) and \lya\ images (bottom panels) for galaxies with optical extent exceeding 20kpc on the side. The data are the same as shown in Figure \ref{fig:lacos-rgb} and \ref{fig:lacos-lya}, but on scales that showcase the emission visible across all filters. The white bar at the bottom of the RGB composite panels indicate the 20 kpc scale. These galaxies all have very tenuous \lya\ emission, which is the reason why it is barely visible in the lower panels.}
    \label{fig:lacos-large_gals}
\end{figure*}

Some of the galaxies in LaCOS have emission on scales extending beyond the 20kpc boxes presented on Figures \ref{fig:lacos-rgb} and \ref{fig:lacos-lya}. Larger cutouts for these galaxies are presented on Figure \ref{fig:lacos-large_gals}.

\end{document}